\documentclass[nohyperref]{article}

\usepackage[utf8]{inputenc}
\usepackage{amsmath}
\usepackage{amssymb}
\usepackage{amsthm}
\usepackage{mathtools}
\usepackage{bm}
\usepackage[round]{natbib}
\usepackage{algorithm}
\usepackage{algpseudocode}
\usepackage[eulergreek]{sansmath}
\usepackage{standalone}
\usepackage{pgfplots}
\usepackage{wrapfig}
\usepackage{microtype}
\usepackage{graphicx}
\usepackage{subfigure}
\usepackage{booktabs} 
\usepackage{hyperref}

\usepackage[accepted]{icml2022}

\usepgfplotslibrary{external}
\usepgfplotslibrary{fillbetween}

\pgfplotsset{compat=1.16}
\pgfplotsset{
  title style={font=\sansmath\sffamily},
  tick label style={font=\sansmath\sffamily},
  every axis label={font=\sansmath\sffamily},
  legend style={font=\sansmath\sffamily},
  label style={font=\sansmath\sffamily},
  every axis plot/.append style={mark=none,line width=1pt},
}

\pgfplotsset{select coords between index/.style 2 args={
    x filter/.code={
        \ifnum\coordindex<#1\def\pgfmathresult{}\fi
        \ifnum\coordindex>#2\def\pgfmathresult{}\fi
    }
}}

\tikzstyle{Harsha}=[red!80!black]
\tikzstyle{MRC}=[blue!50!white]
\tikzstyle{ORC}=[blue!50!black]
\tikzstyle{RS}=[yellow!50!red!90!black]
\tikzstyle{PFR}=[black,mark=o,only marks]
\tikzstyle{PFR2}=[black]
\tikzstyle{hybrid}=[green!50!orange]

\theoremstyle{plain}
\newtheorem{theorem}{Theorem}[section]

\newtheorem{lemma}[theorem]{Lemma}
\newtheorem{corollary}[theorem]{Corollary}
\newcounter{recounter}
\newtheorem{retheorem}{Theorem}[recounter]
\newtheorem{recorollary}{Corollary}[recounter]
\theoremstyle{definition}

\theoremstyle{remark}

\icmltitlerunning{Algorithms for the Communication of Samples}

\DeclarePairedDelimiterX{\infdivx}[2]{[}{]}{%
  #1\;\|\;#2%
}
\DeclarePairedDelimiterX{\infdivy}[2]{[}{]}{%
  #1, #2%
}
\newcommand{\KL}{D_\textnormal{KL}\infdivx}
\newcommand{\TV}{D_\textnormal{TV}\infdivy}

\DeclareMathOperator*{\argmax}{argmax\,}
\DeclareMathOperator*{\argmin}{argmin\,}

\newcommand{\cx}{\mathbf{c}_\mathbf{x}}

\newcommand{\x}{\mathbf{x}}

\newcommand{\z}{\mathbf{z}}
\newcommand{\X}{\mathbf{X}}

\newcommand{\Z}{\mathbf{Z}}
\newcommand{\U}{\mathbf{U}}
\newcommand{\EX}{\mathbb{E}}

\newcommand{\idx}{{N^*}}
\newcommand{\idxis}{{N_\textnormal{MRC}^*}}
\newcommand{\idxrs}{{N_\textnormal{RS}^*}}
\newcommand{\idxsis}{{N_\textnormal{ORC}^*}}
\newcommand{\idxpfr}{{N_\textnormal{PFR}^*}}
\newcommand{\tildeidx}{{\tilde N^*}}
\newcommand{\bks}{\mathbf{K}^*}

\newcommand{\wmin}{w_\textnormal{min}}
\newcommand{\nstar}{{n^*}}

\begin{document}

\twocolumn[
    \icmltitle{Algorithms for the Communication of Samples}
    \icmlsetsymbol{equal}{*}
    
    \begin{icmlauthorlist}
        \icmlauthor{Lucas Theis}{googleuk}
        \icmlauthor{Noureldin Yosri}{googleir}
    \end{icmlauthorlist}
    
    \icmlaffiliation{googleuk}{Google, London, UK}
    \icmlaffiliation{googleir}{Google, Dublin, Ireland}
    
    \icmlcorrespondingauthor{Lucas Theis}{theis@google.com}
    
    \icmlkeywords{channel simulation, reverse channel coding, random coding}
    
    \vskip 0.3in
]

\begin{abstract}
The efficient communication of noisy data has applications in several areas of machine learning, such as neural compression or differential privacy, and is also known as reverse channel coding or the channel simulation problem. Here we propose two new coding schemes with practical advantages over existing approaches. First, we introduce \textit{ordered random coding} (ORC) which uses a simple trick to reduce the coding cost of previous approaches. This scheme further illuminates a connection between schemes based on importance sampling and the so-called \textit{Poisson functional representation}. Second, we describe a hybrid coding scheme which uses dithered quantization to more efficiently communicate samples from distributions with bounded support.
\end{abstract}

\printAffiliationsAndNotice{}

\section{Introduction}

Consider a problem where a sender has information $\x$ and wants to communicate a noisy version of it over a digital channel, 
\begin{align}
    \Z \sim \x + \U.
\end{align}
The sender does not care which value the noise $\U$ takes as long as it follows a given distribution. For example, it may be desired that the noise is a fair sample from a Gaussian distribution. Can we exploit the sender's indifference to the exact value of the noise to save bits in the communication? More generally, we may want to send a sample from a given distribution,
\begin{align}
    \mathbf{Z} \sim q_\x.
\end{align}
How can we communicate such a sample most efficiently? This is the problem of \textit{reverse channel coding}. While channel coding tries to communicate digital information over a noisy channel with as few errors as possible, reverse channel coding attempts to do the opposite, namely to simulate a noisy channel over a digital channel. This problem has therefore also been referred to as ``channel simulation'' \citep[e.g.,][]{cuff2008} and is closely related to ``relative entropy coding'' \citep{flamich2020cwq}.

The reverse channel coding problem occurs in many applications. In neural compression, differentiable channels enable gradient-based methods to optimize encoders but these channels are necessarily noisy if we want to limit their capacity. For example, it is common to approximate quantization with uniform noise when training neural networks for lossy compression \citep{balle2017end}. Reverse channel coding allows us to implement such noisy channels at test time \citep{agustsson2020uq} and to use arbitrary distributions in place of uniform noise \citep{havasi2018miracle}.

In differential privacy, most mechanisms seek to limit the amount of sensitive information revealed to another party by adding noise to the data \citep{dwork2006dp}. Efficiently communicating such private information is an active area of research \citep[e.g.,][]{chen2020trilemma,shah2021dp} and the goal of reverse channel coding.

Quantum teleportation can be viewed as another instance of reverse channel coding where classical bits are used to communicate stochastic information in the form of a qubit. Some of the earliest results on reverse channel coding were obtained in quantum mechanics \citep{bennett2002reverse}.

The naive approach to our problem would be to let the sender generate a sample and then to encode this noisy data. If the data or the noise stems from a continuous distribution, lossless coding is impossible as it would require an infinite number of bits. On the other hand, lossy coding (by first quantizing) leads to further corruption of the data and still wastes bits on encoding the remaining noise. In contrast, efficient reverse channel coding techniques are able to communicate such information with a coding cost which is close to the mutual information between the data $\X$ and the sample $\Z$ \citep[e.g.,][]{li2018pfr},
\begin{align}
    I[\X, \Z] = h[\Z] - h[\Z \mid \X] = \EX[\KL{q_\X}{p}],
    \label{eq:mi}
\end{align}
where $h$ is the differential entropy and $p$ is the marginal distribution of $\Z$. Unlike the naive approach, the coding cost actually decreases as we introduce more noise, that is, when the (differential) entropy of $q_\X$ increases.

More formally, a \textit{reverse channel code} consists of an encoder $f$ and a decoder $g$,
\begin{align}
    f: \mathcal{X} \times [0, 1) \rightarrow \mathbb{N}_0, \quad
    g: \mathbb{N}_0 \times [0, 1) \rightarrow \mathcal{Z},
\end{align}
where $\mathcal{X}$ and $\mathcal{Z}$ are arbitrary spaces. The encoder takes $\mathbf{x} \in \mathcal{X}$ together with a possibly infinite number of bits represented by a real number $u \in [0, 1)$ and outputs a discrete representation $k \in \mathbb{N}_0$. The decoder accepts $k$ and $u$ and outputs $\mathbf{z} \in \mathcal{Z}$.

Assume $\X$ and $\Z$ are random variables jointly distributed over $\mathcal{X} \times \mathcal{Z}$ such that $I[\X, \Z] < \infty$. Let $K = f(\X, U)$ where $U$ is a random variable uniformly distributed over $[0, 1)$ and independent of $\X$. For simplicity, we will assume that for all $\x \in \mathcal{X}$ the conditional distribution of $\Z$ given $\X = \x$ has a density $q_\x$  with respect to a Borel or counting measure on $\mathcal{Z}$. A \textit{reverse channel coding problem} is any problem which tries to find a code such that $H[K \mid U]$ is small and the distribution of $g(K, U)$ is simultaneously close to $q_\x$ in some well-defined sense. In this paper, we will measure closeness in terms of the total variation divergence (Eq.~\ref{eq:tvd}).

Here we assume that the encoder and decoder have access to a \textit{shared source of randomness} $U$ which we may therefore also be used to encode $K$ at a coding cost close to $H[K \mid U]$. Other variants of reverse channel coding limit the amount of shared randomness which can be used  but are not considered here.

In the following, we will first provide an overview of a few useful algorithms implementing reverse channel codes. For instance, we will describe a practical algorithm based on the so-called Poisson functional representation \citep{li2018pfr}.
We will then introduce new algorithms with practical advantages over existing approaches along with theoretical results on their properties, which represents our main contribution. As a further contribution, we will present a unifying view of some of the algorithms which helps to clarify the relationship between them and sheds light on their empirical behavior. Finally, we will provide the first direct empirical comparison between different reverse channel coding algorithms. All proofs and additional empirical results can be found in the appendix.

\section{Related work}

The \textit{reverse Shannon theorem} of \citet{bennett2002reverse} shows that a sender who has access to $\X$ can communicate an instance of $\Z$ at a cost which is close to the two random variables' mutual information. Many papers have considered problems related to reverse channel coding and derived bounds on the coding cost of communicating a sample \citep[e.g.,][]{cover2007capacity,harsha2007,braverman2014}. To our knowledge, the sharpest known upper bound was provided by \citet{li2021lemma} who showed that an optimal code using shared randomness does not require more than
\begin{align}
    I[\X, \Z] + \log (I[\X, \Z] + 1) + 4.732
    \label{eq:bound}
\end{align}
bits on average to communicate an exact sample. On the other hand, \citet{li2018pfr} showed that distributions exist for which the coding cost is at least
\begin{align}
    I[\X, \Z] + \log (I[\X, \Z] + 1) - 1.
    \label{eq:lower_bound}
\end{align}
That is, the bound in Eq.~\ref{eq:bound} cannot be improved significantly without making additional assumptions about the distributions involved \citep[see also][]{braverman2014}. Note that the communication overhead (the second and third term) becomes relatively less important as the transmitted amount of information increases.

Most general reverse channel coding algorithms operate on the same basic principle. First, a potentially large number of candidates is generated from a fixed distribution which is known to the sender and receiver,
\begin{align}
    \Z_n \sim p,
\end{align}
where $n \in \mathbb{N}$ or $n \in \{1, \dots, N\}$. Both the sender and receiver are able to generate these candidates without communication by using a shared source of randomness. 
In practice, this will typically be a pseudorandom number generator with a common seed that has been established in advance.
The sender selects an index $N^*$ according to some distribution such that, at least approximately,
\begin{align}
    \Z_{N^*} \sim q_\x.
\end{align}
Note that only $N^*$ needs to be communicated and this can be done efficiently if $H[N^*]$ is small. The main difference between algorithms is in how $N^*$ is decided.

\citet{li2017dyadic} described an algorithm for communicating samples from distributions with log-concave PDFs without common randomness. Without a shared source of randomness, the number of bits required is at least \textit{Wyner's common information} \citep{wyner1975ci,cuff2008}, which can be significantly larger than the mutual information \citep{xu2011wyner}. In the following, we therefore focus on algorithms with access to common randomness.

\citet{agustsson2020uq} showed that there is no general algorithm whose computational complexity is polynomial in the communication cost. That is, as the amount of information transmitted increases, general purpose algorithms become prohibitively expensive. One solution to this problem is to split information into chunks and to encode these chunks separately \citep{havasi2018miracle,flamich2020cwq}. However, this reduces statistical efficiency as each chunk will contribute its own overhead to the overall coding cost. We therefore typically find tension between the computational efficiency and the coding efficiency of a scheme.

A more well-known idea in machine learning is \textit{bits-back coding} \citep{wallace1990bb,hinton1993bb} which at first glance may appear closely related to reverse channel coding. Here, the goal is to losslessly compress a source~$\X$ using a model of its joint distribution with a set of latent variables~$\Z$. Encoding an instance~$\x$ involves sampling $\Z \sim q_\x$ while using previously encoded bits as a source of randomness. The data and latent variables are subsequently encoded using the model's joint distribution \citep{townsend2019bb}. Unlike reverse channel coding, however, bits-back coding necessarily transmits a perfect copy of the data, that is, it is an implementation of lossless source coding. On the other hand, reverse channel coding can be viewed as a generalization of source coding. Lossless source coding is recovered as a special case when choosing $q_\x(\z) = \delta(\z - \x)$.

\section{Algorithms}

We will first continue the discussion of related work by introducing existing algorithms for the simulation of noisy channels. New methods and results are presented in Sections~\ref{sec:sis}, \ref{sec:unify}, and \ref{sec:hybrid}.

\subsection{Rejection sampling}
\label{sec:rs}

\textit{Rejection sampling} (RS) is a method for generating a sample from one distribution given samples from another distribution. As an introductory example, we show how RS can be turned into a reverse channel coding scheme.

Let $\Z_n$ be candidates drawn independently from a proposal distribution $p$. Further, let $U_n \sim \text{Uniform}([0, 1))$. RS selects the first index $\idxrs$ such that
\begin{align}
    U_\idxrs \leq \wmin \frac{q_\x(\Z_\idxrs)}{p(\Z_\idxrs)}
    \label{eq:rs}
\end{align}
where $\wmin$ is any number such that
\begin{align}
    \wmin \leq \inf_\z \frac{p(\z)}{q_\x(\z)},
    \label{eq:rs_constant}
\end{align}
ensuring that the right-hand side in Eq.~\ref{eq:rs} is smaller than 1. If $\wmin > 0$, then $\idx$ will be finite and, crucially,
\begin{align}
    \Z_\idxrs \sim q_{\x}.
\end{align}
A sender could thus communicate a sample from $q_\x$ by sending $\idxrs$, assuming the receiver already has access to the candidates $\Z_n$. Note that this works even when the distribution is continuous since $\idxrs$ will still be discrete. While $\wmin$ can be chosen to depend on $\x$, in the following analysis we assume for simplicity that the same value is chosen for all target distributions $q_\x$.

Let us consider the coding cost of encoding $\idxrs$. The average probability of accepting a candidate is
\begin{align}
    \int p(\z) \wmin \frac{q_\x(\z)}{p(\z)} \, d\z = \wmin.
\end{align}
The marginal distribution of $\idxrs$ is therefore a geometric distribution whose entropy can be bounded by
\begin{align}
    H[\idxrs] 
    &\leq -\log \wmin + 1/\ln 2.
\end{align}
Rejection sampling is efficient if $H[\idxrs]$ is not much more than the information contained in $\Z$. While it is easy to construct examples where RS is efficient, it is also easy to construct examples where $-\log\wmin$ is significantly larger than $I[\X, \Z]$. For instance, the density ratio in Eq.~\ref{eq:rs_constant} may be unbounded.
However, if we are willing to accept an approximate sample, then there are ways to limit the coding cost even then. For example, we may unconditionally accept the $N$th candidate if the first $N - 1$ candidates are rejected. In this case, the distribution of $\Z_\idx$ will be a mixture distribution with density
\begin{align}
    \beta p(\z) + (1 - \beta) q_\x(\z),
\end{align}
where $\beta = (1 - \wmin)^{N - 1}$ is the probability of rejecting all $N - 1$ candidates.
The quality of a sample is often measured in terms of the \textit{total variation distance} (TVD),
\begin{align}
    \TV{p}{q} = \frac{1}{2} \int |p(\z) - q(\z)| \, d\z.
    \label{eq:tvd}
\end{align}
When measuring the distance of the mixture distribution from the target distribution $q_\x$, we obtain
\begin{align}
    \TV{\beta p + (1 - \beta) q_\x}{q_\x} &= \beta \TV{p}{q_\x}.
\end{align}
That is, the divergence decays exponentially with~$N$.

An alternative approach to limiting the coding cost is to choose an invalid but larger $\wmin$. \citet{harsha2007} described a related approach which effectively uses $\wmin = 1$ but is nevertheless able to produce an exact sample by adjusting the target distribution after each candidate rejection (Appendix~A). However, their approach is computationally expensive and often infeasible for continuous distributions.

\subsection{Minimal random coding}
\label{sec:is}

An approach closely related to \textit{importance sampling} was first considered by \citet{cuff2008} and later dubbed \textit{likelihood encoder} \citep{song2016ld}.
The approach was independently rediscovered in machine learning by \citet{havasi2018miracle} who referred to it as \textit{minimal random coding} (MRC) and used it for model compression. It has since also been used for lossy image compression \citep{flamich2020cwq}. Unlike \citet{havasi2018miracle}, \citet{cuff2008} only considered discrete distributions and assumed that $p$ is the true marginal distribution of the data. But \citet{cuff2008} also described a more general approach where the amount of shared randomness between the sender and receiver is limited.

In MRC, the sender picks one of $N$ candidates by sampling an index $\idxis$ from the distribution
\begin{align}
    \pi_\x(n) \propto q_\x(\Z_n)/p(\Z_n).
    \label{eq:is}
\end{align}
Unlike RS, the distribution of $\Z_\idxis$ (call it $\tilde q_\x$) will in general only approximate $q_\x$. On the other hand, the coding cost of MRC can be significantly smaller. \citet{havasi2018miracle} showed that under reasonable assumptions, samples from $\tilde q_\x$ will be similar to samples from $q_\x$ if the number of candidates is
\begin{align}
    N = 2^{\KL{q_\x}{p} + t}
    \label{eq:is_cc}
\end{align}
for some $t > 0$. 
That is, the number of candidates required to guarantee a sample of high quality grows exponentially with the amount of information gained by the receiver.

Since acceptable candidates may appear anywhere in the sequence of candidates, each index is a priori equally likely to be picked. That is, the marginal distribution of $\idxis$ is uniform and its entropy is
\begin{align}
    H[\idxis] = \log N.
\end{align}

\subsection{Poisson functional representation}
\label{sec:pfr}

\citet{li2018pfr} studied the following \textit{Poisson functional representation} (PFR) of a random variable. Let $T_n$ be the arrival times of a homogeneous Poisson process on the non-negative real line such that $T_n \geq 0$ and $T_n \leq T_{n + 1}$ for all $n \in \mathbb{N}$. Let
\begin{align}
    \Z_n &\sim p, &
    \idxpfr &= \argmin_{n \in \mathbb{N}} T_n \frac{p(\Z_n)}{q_\x(\Z_n)} \label{eq:pfr}.
\end{align}
for all $n \in \mathbb{N}$. Then $\Z_\idxpfr$ has the distribution $q_\x$. The same construction was already considered by \citet{maddison2016ppmc} for the purpose of Monte Carlo integration but not for channel simulation.

As in RS, the index $\idxpfr$ picks one of infinitely many candidates and we obtain an exact sample from the target distribution.
However, \citet{li2018pfr} provided much stronger guarantees for the coding cost of the $\idxpfr$. In particular,
\begin{align}
    H[\idxpfr] \leq I[\X, \Z] + \log(I[\X, \Z] + 1) + 4.
\end{align}
The distribution of $\idxpfr$ takes a more complicated form than in RS or MRC \citep[][Eq. 29]{li2021lemma}. Nevertheless, a coding cost corresponding to the bound above can be achieved by entropy encoding $\idxpfr$ with a simple Zipf distribution $p_\lambda(n) \propto n^{-\lambda}$ where \citep{li2018pfr}
\begin{align}
    \lambda = 1 + 1 / (I[\X, \Z] + e^{-1}\log e + 1).
    \label{eq:lambda}
\end{align}

A downside of the PFR is that it depends on an infinite number of candidates. Unlike rejection sampling, we cannot consider the candidates in sequence but generally have to consider the scores of all candidates (Eq.~\ref{eq:pfr}). However, if we can bound the density ratio as in rejection sampling (Eq.~\ref{eq:rs_constant}), then we can terminate our search for the best candidate after considering a finite number of them. Let
\begin{align}
    S_n^* = \min_{i \leq n} T_i \frac{p(\Z_i)}{q_\x(\Z_i)}
\end{align}
be the smallest score observed after taking into account $n$ candidates. Since $T_{m} \geq T_n$ for all $m > n$, all further scores will be at least $T_n \wmin$. Hence, if $S_n^* \leq T_n w_\text{min}$, we can terminate the search. Algorithm~\ref{alg:pfr} summarizes this idea. Here, \texttt{simulate($n$, p)} is a function which simulates a distribution \texttt{p} by returning the $n$th pseudorandom sample derived from $n$ and an implicit random seed. Similarly, $\texttt{expon}(n, 1)$ simulates an exponential distribution with rate 1.
The computational complexity of this algorithm is the same as RS, with an expected number of iterations of $1/\wmin$ \citep{maddison2016ppmc}.

\begin{algorithm}[t]
  \caption{PFR}
  \label{alg:pfr}
  \begin{algorithmic}[1]
  \Require $\texttt{p}, \texttt{q}, w_\text{min}$
    \State $t, n, s^* \gets 0, 1, \infty$
    \Statex
    \Repeat
      \State $z \gets \texttt{simulate}(n, \texttt{p})$
      \Comment{Candidate generation}
      \State $t \gets t + \texttt{expon}(n, 1)$
      \Comment{Poisson process}
      \State $s \gets t \cdot \texttt{p}(z) / \texttt{q}(z)$
      \Comment{Candidate's score}
      \Statex
      \If{$s < s^*$}
        \Comment{Accept/reject candidate}
        \State $s^*, n^* \gets s, n$
      \EndIf
      \Statex
      \State $n \gets n + 1$
    \Until{$s^* \leq t \cdot w_\text{min}$}
    \Statex
    \State \textbf{return} $n^*$
  \end{algorithmic}
\end{algorithm}

\subsection{Ordered random coding}
\label{sec:sis}

In the following we show that, interestingly, a slight modification of MRC is able to reduce the entropy of the selected index while maintaining the exact distribution of the communicated sample.

To generate a sample from $\pi_\x$ (Eq.~\ref{eq:is}), we can write
\begin{align}
    \idx &= \argmax_{n \leq N}\log q_\x(\Z_n) - \log p(\Z_n) + G_n,
    \label{eq:gumbel_soft_max}
\end{align}
where the $G_n$ are Gumbel distributed random variables \citep{gumbel1954} with scale parameter 1. This is the well-known Gumbel-max trick for sampling from a categorical distribution \citep{maddison2014astar}. We can show that this trick still works if we arbitrarily permute the Gumbel random variables, as long as the permutation does not depend on the values of the candidates $\Z_n$. In particular, we have the following result.
\begin{theorem}
Let $\tilde G_n$ be the result of sorting the Gumbel random variables in decreasing order such that
$\tilde G_1 \geq \dots \geq \tilde G_N$
and define
\begin{align}
    \tildeidx &= \argmax_{n \leq N}\log q_\x(\Z_n) - \log p(\Z_n) + \tilde G_n.
\end{align}
Then $\Z_{\idx} \sim \Z_{\tildeidx}$.
\label{th:orc}
\end{theorem}
While the distribution of $\Z_{\tildeidx}$ remains the same, the distribution of $\tildeidx$ is no longer uniform but biased towards smaller indices. Since $\idx$ is uniform, we must have $H[\tildeidx] \leq H[\idx]$. In the next section, we show that $H[\tildeidx]$ is in fact on a par with $H[\idxpfr]$. We dub the approach \textit{ordered random coding} (ORC) and pseudocode for ORC is provided in Appendix~B.

As for the sample quality, it improves quickly once the coding cost exceeds the information contained in a sample. In particular, we have the following corollary to the results of \citet{chatterjee2018is} and \citet{havasi2018miracle}.
\begin{corollary}
    Let $\tilde q_\x$ be the distribution of $\Z_{\tildeidx}$ where $\tildeidx$ is defined as in Theorem~\ref{th:orc}. If the number of candidates is $N = 2^{\KL{q_\x}{p} + t}$ for some $t \geq 2 e^{-1}\log e$ and $p(\z) / q_\x(\z) \geq \wmin > 0$ for all $\z$, then
    \begin{align}
        \TV{\tilde q_\x}{q_\x} \leq 4\varepsilon
    \end{align}
    where with $B = -\log \wmin$ we have
    \begin{align}
         \epsilon \leq 2^{-t/8} + \sqrt{2} \exp\left( -\frac{1}{4B^2}(t/2 - e^{-1} \log e)^2 \right).
    \end{align}
    \label{th:orc_tvd}
\end{corollary}

\subsection{A unifying view}
\label{sec:unify}

In this section we show a close connection between methods based on importance sampling and the PFR. First, we can rewrite Eq.~\ref{eq:gumbel_soft_max} as follows,
\begin{align}
    \idxis &= \argmin_{n \leq N} S_n \frac{p(\Z_n)}{q_\x(\Z_n)}
\end{align}
where $S_n$ are exponentially distributed with rate $1$. This is true because $-\log S_n$ is Gumbel distributed. Note the similarity to the PFR,
\begin{align}
    \idxpfr &= \argmin_{n \in \mathbb{N}} T_n \frac{p(\Z_n)}{q_\x(\Z_n)},
\end{align}
where $T_n \sim \sum_{m = 1}^n S_m$. ORC, on the other hand, first sorts the exponential random variables, $\tilde S_1 \leq \dots \leq \tilde S_N$, before choosing
\begin{align}
    \idxsis &= \argmin_{n \leq N} \tilde S_n \frac{p(\Z_n)}{q_\x(\Z_n)}.
\end{align}
It turns out that \citep{renyi1953os}
\begin{align}
    \tilde S_n \sim \sum_{m = 1}^n \frac{S_m}{N - m + 1}.
    \label{eq:sorted_exponential}
\end{align}
This allows us to generate the sorted exponential variables in $O(N)$ instead of $O(N \log N)$ time with sorting. More interestingly, Eq.~\ref{eq:sorted_exponential} reveals a close connection to the PFR. Where the PFR uses cumulative sums of exponential random variables, ORC uses weighted sums. This representation allows us to arrive at the following result.
\vspace{6pt}
\begin{theorem}
    Let $S_n$ be exponentially distributed RVs and $\Z_n \sim p$ for all $n \in \mathbb{N}$ (i.i.d.), and let 
    \begin{align}
        T_n &= \sum_{m = 1}^n S_m, &
        \tilde T_{N,n} &= \sum_{m = 1}^n \frac{N}{N - m + 1} S_m
    \end{align}
    for $N \in \mathbb{N}$. Let
    \begin{align}
        \idxpfr &= \argmin_{n \in \mathbb{N}} T_n \frac{p(\Z_n)}{q_\x(\Z_n)}, \\
        \idxsis &= \argmin_{n \leq N} \tilde T_{N,n} \frac{p(\Z_n)}{q_\x(\Z_n)}.
    \end{align}
    Then $\idxsis \leq \idxpfr$. Further, if $\idxpfr$ is finite then there exists an $M \in \mathbb{N}$ such that for all $N \geq M$ we have $\idxsis = \idxpfr$.
    \label{th:orc_pfr}
\end{theorem}
 Note that the additional factor $N$ in the definition of $\tilde T_{N,n}$ compared to $\tilde S_n$ does not change the output of argmin. Using Theorem~\ref{th:orc_pfr}, it is not difficult to see that the bound on the coding cost of the PFR also applies to ORC, or the following result.
\vspace{6pt} 
\begin{corollary}
    Let $C = \mathbb{E}_\X[\KL{q_\X}{p}]$ and let $\idxsis$ be defined as in Theorem~\ref{th:orc_pfr}. Then
    \begin{align}
        H[\idxsis] < C + \log (C + 1) + 4.
    \end{align}
    \label{th:orc_entropy}
\end{corollary}
\vspace{-12pt}
To achieve this bound, a Zipf distribution $p_\lambda(n) \propto n^{-\lambda}$ with $\lambda = 1 + 1/(C + e^{-1}\log e + 1)$ can be used to entropy encode the index, analogous to the PFR (Eq.~\ref{eq:lambda}).

The significance of these results is as follows. \citet{li2018pfr} showed that the PFR is near-optimal in the sense that the entropy of $\idxpfr$ is close to the worst-case coding cost needed for communicating a perfect sample. However, the construction of the PFR relies on an infinite number of candidates and there are no theoretical guarantees of the sample quality if we naively limit the number of candidates to $N$. In particular, we do not know how quickly the quality of a communicated sample deteriorates as we decrease~$N$, or how large $N$ should be.
On the other hand, we do have some idea of the quality of a sample obtained via importance sampling from a finite number of candidates \citep[e.g., Corollary~\ref{th:orc_tvd} or the results of][]{cuff2008,chatterjee2018is,havasi2018miracle}. But the coding cost of MRC is relatively large and continues to grow unbounded as $N$ increases. ORC combines the best of both by inheriting the bounds on the coding cost of the PFR and the sample quality of MRC.

Unlike the PFR, the guarantees of ORC hold for a finite number of samples. Unlike MRC, we can make $N$ arbitrarily large without having to worry about an exploding coding cost, which makes it easier to tune this parameter.

\subsection{Dithered quantization}
\label{sec:dq}

\textit{Dithered quantization}, also known as \textit{universal quantization} \citep{ziv1985universal}, refers to quantization with a randomly shif\-ted lattice. Consider a scalar $y \in \mathbb{R}$ and a random variable $U$ uniform over any interval of length one, such as $[0, 1)$. Then \citep[e.g.,][]{schuchman1964dither}
\begin{align}
    \lfloor y - U \rceil + U \sim y + U_0,
\end{align}
where $U_0$ is uniform over $[-0.5, 0.5)$. More generally, let $Q$ be a quantizer which maps inputs to the nearest point on a lattice and let $\mathbf{V}$ be a random variable which is uniformly distributed over an arbitrarily placed Voronoi cell of the lattice. Further, let $\mathbf{V}_0$ be uniform over the Voronoi cell which contains the lattice point at zero. Then \citep{zamir2014book}
\begin{align}
    Q(\mathbf{y} - \mathbf{V}) + \mathbf{V} \sim \mathbf{y} + \mathbf{V}_0.
\end{align}

Dithered quantization has been mainly used as a tool for studying quantization from a theoretical perspective. However, already \citet{roberts1962noise} considered some of its practical advantages over uniform quantization for compressing grayscale images, especially in terms of perceptual quality. \citet{theis2021stochastic} showed that universal quantization can outperform vector quantization where a realism constraint is considered. Universal quantization also recently started being used in neural compression \citep{choi2019uq}, in particular to realize differentiable training losses at inference time \citep{agustsson2020uq}. 

To communicate a sample of a uniform distribution centered around $y$, the sender encodes $K = \lfloor y - U \rceil$.
The receiver decodes $K$ and computes $Z = K + U$, which is distributed as $y + U_0$. Like the reverse channel coding schemes discussed so far, this requires a shared source of randomness in the form of $U$. \citet{zamir1992universal} showed that dithered quantization is statistically efficient in the sense that
\begin{align}
    H[K \mid U] = I[Y, Z].
\end{align}
That is, the cost of encoding $K$ is as close as can be to the amount of information contained in $Z$. Note that we can condition on $U$ when encoding $K$ since $U$ is known to both sender and receiver. Another advantage of dithered quantization is that it is computationally highly efficient, at least for the simple case of the integer lattice.

\subsection{Hybrid coding}
\label{sec:hybrid}

\begin{algorithm}[t]
  \caption{Hybrid coding}
  \label{alg:hybrid}
  \begin{algorithmic}[1]
  \Require $c, \texttt{q}, w_\text{min}, N$
    \State $t, n, s^* \gets 0, 1, \infty$
    \Statex
    \Repeat
      \State $u \gets \texttt{uniform}(n, 0, 1)$
      \Comment{Candidate generation}
      \State $k \gets \texttt{round}(c - u)$
      \State $z \gets k + u$
      \Statex
      \State $v \gets N / (N - n + 1)$
      \State $t \gets t + v \cdot \texttt{expon}(n, 1)$
      \State $s \gets t / \texttt{q}(z)$
      \Comment{Candidate's score}
      \Statex
      \If{$s < s^*$}
        \Comment{Accept/reject candidate}
        \State $s^*, n^*, k^* \gets s, n, k$
      \EndIf
      \Statex
      \State $n \gets n + 1$
    \Until{$s^* \leq t \cdot w_\text{min}$ \textbf{or} $n > N$}
    \Statex
    \State \textbf{return} $n^*, k^*$
  \end{algorithmic}
\end{algorithm}

While dithered quantization is computationally much more efficient than general reverse channel coding schemes, it is also much more limited in terms of the distributions it can simulate. Here we propose a hybrid coding scheme for continuous distributions which retains most of the flexibility of general purpose schemes but is computationally more efficient when the support of the target distribution is small.

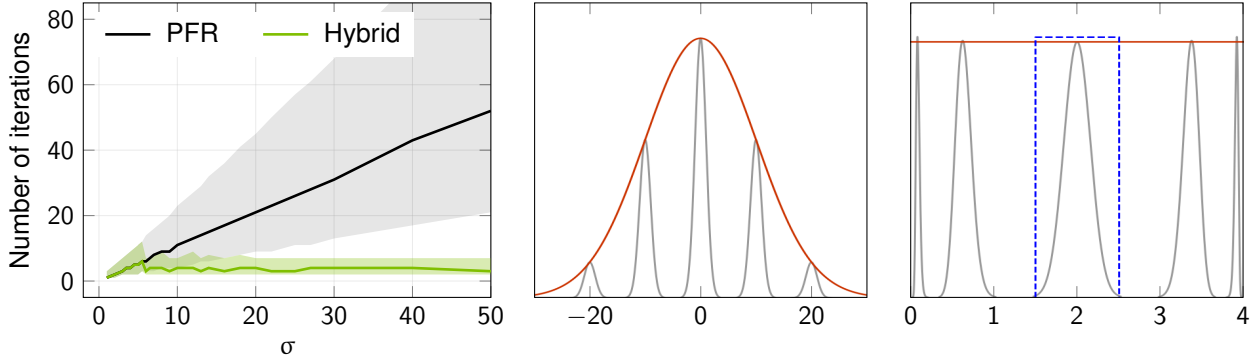
\begin{figure*}[t]
    \centering
        \begin{tikzpicture}
        \begin{axis}[
                xlabel={$\sigma$},
                ylabel={Number of iterations},
                width=7cm,
                height=5.5cm,
                legend columns=-1,
                legend pos=north west,
                legend style={
                    draw=none,
                    /tikz/every even column/.append style={column sep=0.5cm},
                    /tikz/every odd column/.append style={column sep=0.12cm}
                },
                ymin=-5,
                ymax=85,
                xmin=-2,
                xmax=50,
                xtick={0,10,20,30,40,50},
                grid,
                minor grid style={opacity=0.3},
                major grid style={opacity=0.3},
            ]

            \addplot[PFR2] table[x=sigma,y=p50] {figures/data/gaussian_1_pfr.dat};
            \addlegendentry{PFR};
            \addplot[hybrid] table[x=sigma,y=p50] {figures/data/gaussian_1_hybrid.dat};
            \addlegendentry{Hybrid};
            
            \addplot[opacity=0,name path=pfr75] table[x=sigma,y=p75] {figures/data/gaussian_1_pfr.dat};
            \addplot[opacity=0,name path=pfr25] table[x=sigma,y=p25] {figures/data/gaussian_1_pfr.dat};
            \addplot[PFR2,opacity=0.1] fill between[of=pfr25 and pfr75];
            
            \addplot[opacity=0,name path=hybrid75] table[x=sigma,y=p75] {figures/data/gaussian_1_hybrid.dat};
            \addplot[opacity=0,dashed,name path=hybrid25] table[x=sigma,y=p25] {figures/data/gaussian_1_hybrid.dat};
            \addplot[hybrid,opacity=0.3] fill between[of=hybrid25 and hybrid75];
        \end{axis}
        
        \begin{axis}[
                xshift=6cm,
                width=6cm,
                height=5.5cm,
                xmin=-30,
                xmax=30,
                ymin=0,
                ymajorticks=false,
                xmajorticks=true,
                axis on top,
            ]
            \addplot graphics[xmin=-30,xmax=30,ymin=0,ymax=1] {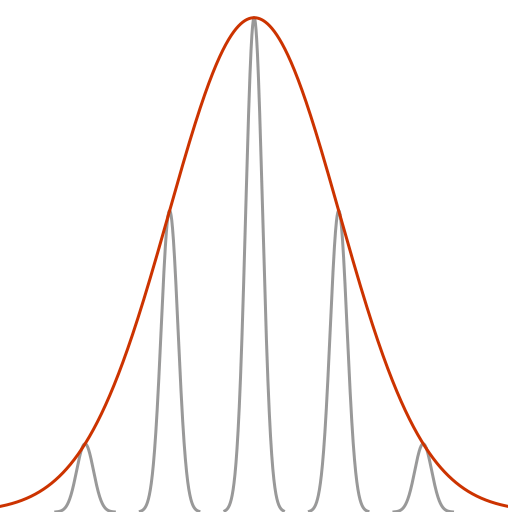};
        \end{axis}
        
        \begin{axis}[
                xshift=11cm,
                width=6cm,
                height=5.5cm,
                xmin=0,
                xmax=4,
                ymin=0,
                ymajorticks=false,
                xmajorticks=true,
                axis on top,
            ]
            \addplot graphics[xmin=0,xmax=4,ymin=0,ymax=1] {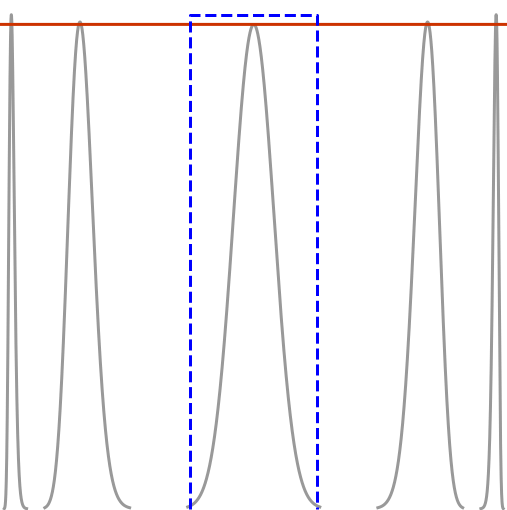};
        \end{axis}
    \end{tikzpicture}
    \vspace{-.2cm}
    \caption{\textit{Left:} Computational cost of communicating a sample from a Gaussian whose mean varies with standard deviation $\sigma$. Solid lines indicate the median number of candidates considered by an algorithm while shaded regions indicate the 25th and 75th percentile. The initial rise and sharp drop of the green curve is due to the discontinuity of $M$ as a function of $\sigma$.
    \textit{Middle:} Illustration of example target distributions $\tilde q_\x$ (gray; scaled for visualization purposes) and the marginal distribution $\mathbb{E}[\tilde q_\x]$ (red). \textit{Right:} The same distributions after transformation with the marginal's CDF, $\Phi_{\sigma^2 + 1}$, and scaling by $M = 4$. The dashed blue line indicates $r_\x$.}
    \label{fig:hybrid}
\end{figure*}

The general idea is as follows. Instead of drawing candidates from a fixed distribution $p$, candidates $\Z_n$ are drawn from a distribution $r_\x$ which acts as a bridge and more closely resembles the target distribution $q_\x$. Since $r_\x$ is closer to $q_\x$, we will require fewer candidates to find one that is suitable. Let 
\begin{align}
    \idx &= \argmin_{n \leq N} \tilde T_{N,n} \frac{r_\x(\Z_n)}{q_\x(\Z_n)},
    \label{eq:idx_hybrid}
\end{align}
be the index of the selected candidate. Unlike before, only the sender has access to the candidates and so knowing $\idx$ alone will not allow us to reconstruct $\Z_\idx$. Our hybrid coding scheme relies on two insights. First, the receiver does not require access to all candidates but only to the selected candidate. Second, the missing information can be encoded easily and efficiently if $r_\x$ can be simulated with dithered quantization.

We assume for now that there exist vectors $\cx$ such that the support of $q_\x$ is contained in the support of
\begin{align}
    r_\x(\z) = \begin{cases}
        1 & \text{ if } \z \in \cx + [-0.5, 0.5)^D, \\
        0 & \text{ else }
    \end{cases}
\end{align}
which can be simulated via dithered quantization,
\begin{align}
    \mathbf{K}_n &= \lfloor \cx - \mathbf{U}_n \rceil, &
    \mathbf{Z}_n &= \mathbf{K}_n + \mathbf{U}_n,
    \label{eq:bks_hybrid}
\end{align}
where $\mathbf{U}_n \sim \text{Uniform}([0, 1)^D)$. Define $\bks = \mathbf{K}_{N^*}$.
Hybrid coding transmits the pair ($\idx, \mathbf{K}^*$), which the receiver uses to reconstruct the selected candidate via
\begin{align}
    \mathbf{Z}_\idx = \mathbf{K}^* + \mathbf{U}_\idx.
\end{align}
\begin{theorem}
    Let $\idx$ and $\bks$ be defined as in Eq.~\ref{eq:idx_hybrid} and below Eq.~\ref{eq:bks_hybrid} and let $p$ be the uniform distribution over $[0, M_1) \times \cdots \times [0, M_D)$ for some $M_i \in \mathbb{N}$. Then
    \begin{align*}
        H[\idx, \bks]
        < C + \log (C - \textstyle\sum_i \log M_i + 1) + 4,
    \end{align*}
    where $C = \mathbb{E}_\X[\KL{q_\X}{p}]$.
    \label{th:hybrid}
\end{theorem}
Theorem~\ref{th:hybrid} shows that $(\idx, \bks)$ is an efficient representation if the marginal distribution of $\Z$ is uniform over some box whose sides have lengths $M_i$, since then $C = I[\X, \Z]$. However, for continuous random variables this can always be achieved through a transformation $\Psi$. If $\tilde q_\x$ is the desired target distribution before the transformation, then
\begin{align}
    q_\x(\z) &= \tilde q_\x(\Psi(\z) ) |D\Psi(\z)|
\end{align}
is the target distribution in transformed space. Note that after the transformation, the support of $q_\x$ is always bounded. Moreover, for small enough $M_i$ the support will be contained in the support of $r_\x$, satisfying our earlier assumption. This is illustrated for Gaussian distributions in Fig~\ref{fig:hybrid}.

\begin{figure*}[t]
    \centering
        \begin{tikzpicture}
        \begin{axis}[
                xlabel={Coding cost [bit]},
                ylabel={Divergence [$D_\text{TV}$]},
                width=5.5cm,
                height=5.5cm,
                legend columns=-1,
                legend style={
                    at={(14.5cm, 4.8cm)},
                    draw=none,
                    /tikz/every even column/.append style={column sep=0.5cm},
                    /tikz/every odd column/.append style={column sep=0.12cm}
                },
                xmin=7,
                xmax=17,
                ymin=-0.05,
                ymax=0.55,
                xtick={0,2,...,16},
                ytick={0.0,0.1,0.2,0.3,0.4,0.5},
                grid,
                minor grid style={opacity=0.3},
                major grid style={opacity=0.3},
            ]
            \addplot[PFR] table[x=cc,y=TV,select coords between index={17}{17}] {figures/data/PFR.dat};
            \addlegendentry{PFR};
            
            \addplot[MRC] table[x=cc,y=TV] {figures/data/MRC.dat};
            \addlegendentry{MRC};
            
           \addplot[ORC] table[x=cc,y=TV] {figures/data/ORC.dat};
           \addlegendentry{ORC};
            
            \addplot[RS] table[x=cc,y=TV] {figures/data/RS.dat};
            \addlegendentry{RS};
            
           \addplot[Harsha] table[x=cc,y=TV] {figures/data/Harsha.dat};
           \addlegendentry{RS*};
            
            \addplot[black!40,densely dashed] coordinates {
                (10.881777,-0.05)
                (10.881777,1)
            };
        \end{axis}
        \begin{axis}[
                xmode=log,
                log basis x={2},
                xshift=11cm,
                xlabel={Number of candidates [$N$]},
                ylabel={Coding cost [bit]},
                xtick={1,8,64,512,4096,32768},
                ytick={0,2,...,16},
                ymin=-1,
                ymax=17,
                xmin=0.5,
                xmax=85536,
                grid,
                minor grid style={opacity=0.3},
                major grid style={opacity=0.3},
                width=5.5cm,
                height=5.5cm,
            ]
            \addplot[RS] table[x=N,y=cc] {figures/data/RS.dat};
            \addplot[Harsha] table[x=N,y=cc] {figures/data/Harsha.dat};
            \addplot[MRC] table[x=N,y=cc] {figures/data/MRC.dat};
            \addplot[ORC] table[x=N,y=cc] {figures/data/ORC.dat};
            \addplot[black!40,densely dashed] coordinates {
                (0.5,10.881777)
                (131072,10.881777)
            };
        \end{axis}
        \begin{axis}[
                xmode=log,
                log basis x={2},
                log basis y={2},
                xshift=5.5cm,
                ylabel={Divergence [$D_\text{TV}$]},
                xlabel={Number of iterations},
                xtick={256,512,1024,2048,4096,8192,16384,32768,65536},
                xticklabels={$2^8$,,$2^{10}$,,$2^{12}$,,$2^{14}$,,$2^{16}$},
                xmin=206,
                xmax=85536,
                ymin=-0.05,
                ymax=0.55,
                ytick={0.0,0.1,0.2,0.3,0.4,0.5},
                grid=both,,
                minor grid style={opacity=0.3},
                major grid style={opacity=0.3},
                width=5.5cm,
                height=5.5cm,
                scaled x ticks=false,
                scaled y ticks=false,
            ]
            \addplot[RS] table[y=TV,x=Imean] {figures/data/RS.dat};
            \addplot[Harsha] table[y=TV,x=Imean] {figures/data/Harsha.dat};
            \addplot[MRC] table[y=TV,x=I50] {figures/data/MRC.dat};
            \addplot[ORC] table[y=TV,x=Imean] {figures/data/ORC.dat};
            \addplot[PFR] table[y=TV,x=Imean,select coords between index={17}{17}] {figures/data/PFR.dat};
        \end{axis}
    \end{tikzpicture}
    \caption{A comparison of various reverse channel coding algorithms for $2^{16}$-dimensional categorical distributions which are themselves randomly distributed according to a Dirichlet distribution. Dashed lines indicate the average KLD between the target distribution and the uniform candidate generating distribution. \textit{Left:} The sample quality as a function of the coding cost. \textit{Middle:} The sample quality as a function of the computational cost (for which the number of iterations is a proxy, except for RS*). \textit{Right:} The coding cost as a function of the maximum number of candidates considered.}
    \label{fig:comparison}
\end{figure*}
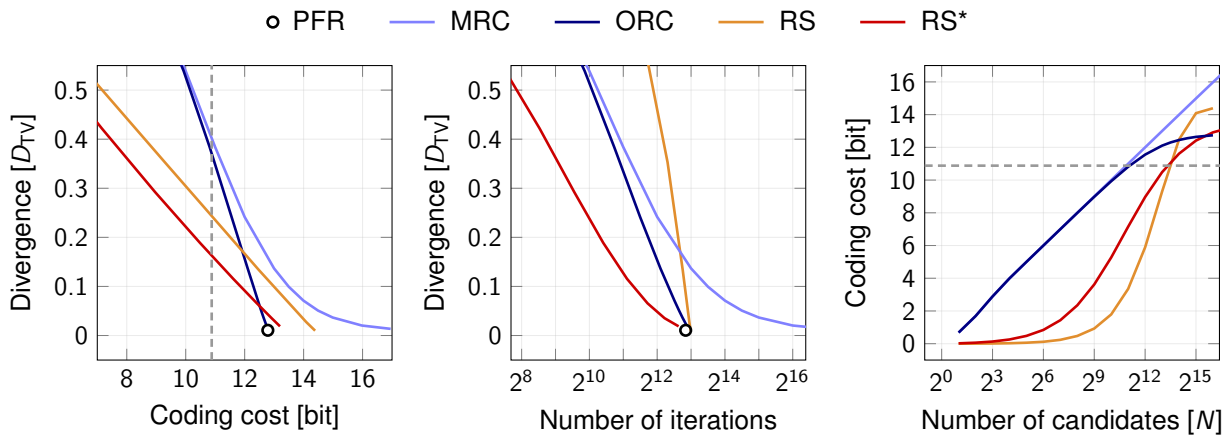

To achieve the bound suggested by Theorem~\ref{th:hybrid}, the sender can first encode $\idx$
\citep[e.g., using arithmetic coding;][]{rissanen1979arithmetic} 
while assuming a Zipf distribution $p_\lambda(n) \propto n^{-\lambda}$ with
\begin{align}
    \textstyle\lambda = 1 + 1 / (C - \sum_i \log M_i + e^{-1} \log e + 1).
\end{align}
The $K_i^*$ are subsequently added to the bit stream using a fixed rate of $\log M_i$ bits.

Algorithm~\ref{alg:hybrid} describes how $N^*$ and $K^*$ are found during the encoding process. Here, \texttt{q} is the transformed density and \texttt{p} is not needed as it is assumed to be uniform. Note that hybrid coding effectively reduces to ORC when the support is unconstrained ($M_i = 1$). For larger $M_i$, the bound on the coding cost improves only slightly but the computational cost reduces significantly. Since the number of candidates required for a sample of high quality grows exponentially in the KLD (Eq.~\ref{eq:is_cc}) and
\begin{align}
    D_\text{KL}[q_\x \mid\mid r_\x] = D_\text{KL}[q_\x \mid\mid p] - \log M_i,
\end{align}
we can expect a speedup on the order of $\prod_i M_i$. We thus want to maximize $M_i$ while making sure that the support of $q_\x$ is still contained within that of $r_\x$.


\section{Experiments}
\label{sec:experiments}

We run two sets of empirical experiments to compare the reverse channel coding schemes discussed above. We first investigate the effect of hybrid coding on the computational cost of communicating a (truncated) Gaussian sample. We then compare the performance of a wider set of algorithms for the task of approximately simulating a categorical distribution. 

\subsection{Gaussian distribution}

Consider the task of communicating a sample from a $D$-dimensional Gaussian with a random mean,
\begin{align}
    \Z &\sim \mathcal{N}(\X, \mathbf{I}), &
    \X &\sim \mathcal{N}(0, \sigma^2 \mathbf{I}),
\end{align}
where $\mathbf{I}$ is the identity matrix and 
the mean $\X$ itself is Gaussian distributed with covariance $\sigma^2\mathbf{I}$. The marginal distribution of $\Z$ is Gaussian with mean zero and covariance $\sigma^2\mathbf{I} + \mathbf{I}$ and so we use this distribution as our candidate generating distribution $p$. The average information gained by obtaining a sample is
\begin{align}
    I[\X, \Z] &= \textstyle -\frac{D}{2} \log\left( 1 - \frac{\sigma^2}{\sigma^2 + 1} \right).
\end{align}
To be able to apply the hybrid coding scheme, we slightly truncate the target distribution by assigning zero density to a small fraction $\theta$ of values with the lowest density. The TVD between the truncated Gaussian and the Gaussian distribution is $\theta$. A classifier observing $\Z$ would be able to distinguish between these two distributions with an accuracy of at most $1/2 + \theta/2$. In our experiments, we fix $\theta = 10^{-4}$ so that this accuracy is close to chance.

We compare hybrid coding (Algorithm~\ref{alg:hybrid}) to ORC with $N = \infty$, which reduces to the PFR (Algorithm~\ref{alg:pfr}). Using an unlimited number of candidates allows us to avoid any further approximations and to focus on the computational cost. Appropriate values for $w_\text{min}$ and $M$ are provided in Appendix~H.

Figure~\ref{fig:hybrid} shows the average number of iterations an algorithm runs before identifying a suitable candidate of a 1-dimensional (truncated) Gaussian. The computational cost of the PFR grows exponentially with the amount of information transmitted, which is approximately $\log\sigma$. On the other hand, the computational cost of the hybrid coding scheme quickly saturates and remains low throughout, allowing for much quicker communication of the Gaussian sample. Results for higher-dimensional Gaussians are provided in Appendix~\ref{sec:additional}.

\subsection{Categorical distribution}

As another example we consider $D$-dimensional categorical distributions distributed according to a Dirichlet distribution with concentration parameter $\bm{\alpha}$. We chose $D = 2^{16}$ and $\alpha_i = 3\cdot 10^{-4}$ for all $i$, leading to sparse target distributions and a uniform marginal distribution. We include rejection sampling (RS) with an optimal choice for $w_\text{min}$ (a different value for each distribution) as well as the greedy rejection sampler (RS*) of \citet{harsha2007} in the comparison. For each method and target distribution, we simulate $10^5$ samples and measure the TVD between the resulting histogram and the target distribution. As a measure of the coding cost, we estimate the entropy of the index distribution obtained by averaging index histograms of 20 different target distributions.

We explore the effects of  limiting the number of candidates available to an algorithm. We find that the sample quality of all methods deteriorates quickly as the coding cost drops below the information contained in exact samples (Fig.~\ref{fig:comparison}, left). RS performed well in the bit-rate constrained regime but not as well when constraining computational cost (Fig.~\ref{fig:comparison}, middle). RS* performed very well in the low bit-rate regime but we point out that its computational complexity is larger by a factor $D$ per iteration compared to the other methods. PFR and ORC performed best for samples of high quality.

Although MRC is the method which is currently the most widely used in machine learning \citep[e.g.,][]{havasi2018miracle,flamich2020cwq,shah2021dp}, we find that here it performs worse than the other methods, mostly due to its coding and computational cost growing unboundedly with the number of candidates (Fig.~\ref{fig:comparison}, right). ORC addresses this issue such that its coding cost converges to that of the PFR.

\section{Discussion}
We demonstrated a close connection between minimal random coding \citep[MRC;][]{havasi2018miracle}, or likelihood encoding \citep{cuff2008,song2016ld}, and the Poisson functional representation \citep[PFR;][]{li2018pfr}. We introduced ordered random coding (ORC), which occupies a space between the two and benefits from the theoretical guarantees of both. In practice, we found that ORC can significantly outperform MRC, especially as the desired sample quality increases (achieving a 20\% coding cost reduction at a TVD of 0.02).

Our second coding scheme enables much more efficient communication of samples from distributions with concentrated support. When the target distributions' support is unbounded, hybrid coding may still be applied after truncation, as in the Gaussian example. While the hybrid scheme is much more efficient than other approaches for the Gaussian example, its cost still grows exponentially with dimensionality. A potential solution may be to generate candidates using more sophisticated lattices than the integer lattice considered here \citep[e.g.,][]{leech1967,zamir2014book}. The Gaussian channel is ubiquitous in machine learning and plays an important role in, for instance, variational autoencoders \citep{kingma2014vae} and differential privacy \citep{dwork2006dp}.
An important task for future research is to characterize other distributions which can be simulated efficiently and which are therefore of great relevance for practical applications.

\subsubsection*{Acknowledgements}

We would like to thank Abhin Shah, Johannes Ballé, Eirikur Agustsson, and Aaron B. Wagner for helpful discussions of the ideas presented in this manuscript.

\bibliographystyle{icml2022}
\bibliography{references}

\begin{thebibliography}{37}
\providecommand{\natexlab}[1]{#1}
\providecommand{\url}[1]{\texttt{#1}}
\expandafter\ifx\csname urlstyle\endcsname\relax
  \providecommand{\doi}[1]{doi: #1}\else
  \providecommand{\doi}{doi: \begingroup \urlstyle{rm}\Url}\fi

\bibitem[Agustsson \& Theis(2020)Agustsson and Theis]{agustsson2020uq}
Agustsson, E. and Theis, L.
\newblock {Universally Quantized Neural Compression}.
\newblock In \emph{Advances in Neural Information Processing Systems 33}, 2020.

\bibitem[Ball{\'e} et~al.(2017)Ball{\'e}, Laparra, and
  Simoncelli]{balle2017end}
Ball{\'e}, J., Laparra, V., and Simoncelli, E.~P.
\newblock {End-to-end Optimized Image Compression}.
\newblock In \emph{International Conference on Learning Representations}, 2017.

\bibitem[Bennett \& Shor(2002)Bennett and Shor]{bennett2002reverse}
Bennett, C.~H. and Shor, P.~W.
\newblock {Entanglement-Assisted Capacity of a Quantum Channel and the Reverse
  Shannon Theorem}.
\newblock \emph{IEEE Trans. Info. Theory}, 48\penalty0 (10), 2002.

\bibitem[Braverman \& Garg(2014)Braverman and Garg]{braverman2014}
Braverman, M. and Garg, A.
\newblock Public vs private coin in bounded-round information.
\newblock In Esparza, J., Fraigniaud, P., Husfeldt, T., and Koutsoupias, E.
  (eds.), \emph{Automata, Languages, and Programming}, pp.\  502--513.
  Springer, 2014.

\bibitem[Chatterjee \& Diaconis(2018)Chatterjee and Diaconis]{chatterjee2018is}
Chatterjee, S. and Diaconis, P.
\newblock The sample size required in importance sampling.
\newblock \emph{The Annals of Applied Probability}, 28\penalty0 (2):\penalty0
  1099--1135, 2018.

\bibitem[Chen et~al.(2020)Chen, Kairouz, and Ozgur]{chen2020trilemma}
Chen, W.-N., Kairouz, P., and Ozgur, A.
\newblock Breaking the communication-privacy-accuracy trilemma.
\newblock In Larochelle, H., Ranzato, M., Hadsell, R., Balcan, M.~F., and Lin,
  H. (eds.), \emph{Advances in Neural Information Processing Systems},
  volume~33, pp.\  3312--3324. Curran Associates, Inc., 2020.

\bibitem[Choi et~al.(2019)Choi, El-Khamy, and Lee]{choi2019uq}
Choi, Y., El-Khamy, M., and Lee, J.
\newblock Variable rate deep image compression with a conditional autoencoder.
\newblock In \emph{Proceedings of the IEEE International Conference on Computer
  Vision}, 2019.

\bibitem[Cover \& Permuter(2007)Cover and Permuter]{cover2007capacity}
Cover, T.~M. and Permuter, H.~H.
\newblock Capacity of coordinated actions.
\newblock In \emph{2007 IEEE International Symposium on Information Theory},
  pp.\  2701--2705, 2007.
\newblock \doi{10.1109/ISIT.2007.4557184}.

\bibitem[Cuff(2008)]{cuff2008}
Cuff, P.
\newblock Communication requirements for generating correlated random
  variables.
\newblock In \emph{2008 IEEE International Symposium on Information Theory},
  pp.\  1393--1397, 2008.

\bibitem[Dwork et~al.(2006)Dwork, McSherry, Nissim, and Smith]{dwork2006dp}
Dwork, C., McSherry, F., Nissim, K., and Smith, A.
\newblock Calibrating noise to sensitivity in private data analysis.
\newblock In Halevi, S. and Rabin, T. (eds.), \emph{Theory of Cryptography},
  pp.\  265--284. Springer Berlin Heidelberg, 2006.

\bibitem[Flamich et~al.(2020)Flamich, Havasi, and
  Hern{\'a}ndez-Lobato]{flamich2020cwq}
Flamich, G., Havasi, M., and Hern{\'a}ndez-Lobato, J.~M.
\newblock {Compressing Images by Encoding Their Latent Representations with
  Relative Entropy Coding}, 2020.
\newblock Advances in Neural Information Processing Systems 34.

\bibitem[Gumbel(1954)]{gumbel1954}
Gumbel, E.~J.
\newblock {Statistical Theory of Extreme Values and Some Practical
  Applications}.
\newblock \emph{U.S. Department of Commerce, National Bureau of Standards}, 33,
  1954.

\bibitem[{Harsha} et~al.(2007){Harsha}, {Jain}, {McAllester}, and
  {Radhakrishnan}]{harsha2007}
{Harsha}, P., {Jain}, R., {McAllester}, D., and {Radhakrishnan}, J.
\newblock {The Communication Complexity of Correlation}.
\newblock In \emph{Twenty-Second Annual IEEE Conference on Computational
  Complexity}, pp.\  10--23, 2007.

\bibitem[Havasi et~al.(2019)Havasi, Peharz, and
  Hernández-Lobato]{havasi2018miracle}
Havasi, M., Peharz, R., and Hernández-Lobato, J.~M.
\newblock {Minimal Random Code Learning: Getting Bits Back from Compressed
  Model Parameters}.
\newblock In \emph{International Conference on Learning Representations}, 2019.

\bibitem[Hinton \& Van~Camp(1993)Hinton and Van~Camp]{hinton1993bb}
Hinton, G.~E. and Van~Camp, D.
\newblock Keeping the neural networks simple by minimizing the description
  length of the weights.
\newblock In \emph{Proceedings of the sixth annual conference on Computational
  learning theory}, pp.\  5--13, 1993.

\bibitem[Kingma \& Welling(2014)Kingma and Welling]{kingma2014vae}
Kingma, D. and Welling, M.
\newblock {Auto-encoding variational Bayes}.
\newblock In \emph{International Conference on Learning Representations}, 2014.

\bibitem[Leech(1967)]{leech1967}
Leech, J.
\newblock Notes on sphere packings.
\newblock \emph{Canadian Journal of Mathematics}, 19:\penalty0 251–267, 1967.

\bibitem[Li \& Anantharam(2021)Li and Anantharam]{li2021lemma}
Li, C.~T. and Anantharam, V.
\newblock A unified framework for one-shot achievability via the poisson
  matching lemma.
\newblock \emph{IEEE Transactions on Information Theory}, 67\penalty0
  (5):\penalty0 2624--2651, 2021.
\newblock \doi{10.1109/TIT.2021.3058842}.

\bibitem[Li \& El~Gamal(2017)Li and El~Gamal]{li2017dyadic}
Li, C.~T. and El~Gamal, A.
\newblock Distributed simulation of continuous random variables.
\newblock \emph{IEEE Transactions on Information Theory}, 63\penalty0
  (10):\penalty0 6329--6343, 2017.
\newblock \doi{10.1109/TIT.2017.2735438}.

\bibitem[{Li} \& {El Gamal}(2018){Li} and {El Gamal}]{li2018pfr}
{Li}, C.~T. and {El Gamal}, A.
\newblock {Strong Functional Representation Lemma and Applications to Coding
  Theorems}.
\newblock \emph{IEEE Transactions on Information Theory}, 64\penalty0
  (11):\penalty0 6967--6978, 2018.
\newblock \doi{10.1109/TIT.2018.2865570}.

\bibitem[Maddison(2016)]{maddison2016ppmc}
Maddison, C.~J.
\newblock {A Poisson process model for Monte Carlo}.
\newblock In Hazan, T., Papandreou, G., and Tarlow, D. (eds.),
  \emph{Perturbation, Optimization, and Statistics}. MIT Press, 2016.

\bibitem[Maddison et~al.(2014)Maddison, Tarlow, and Minka]{maddison2014astar}
Maddison, C.~J., Tarlow, D., and Minka, T.
\newblock {$A\ast$ Sampling}.
\newblock In \emph{Advances in Neural Information Processing Systems},
  volume~27, 2014.

\bibitem[Rissanen \& Langdon(1979)Rissanen and Langdon]{rissanen1979arithmetic}
Rissanen, J. and Langdon, G.~G.
\newblock Arithmetic coding.
\newblock \emph{IBM Journal of research and development}, 23\penalty0
  (2):\penalty0 149--162, 1979.

\bibitem[Roberts(1962)]{roberts1962noise}
Roberts, L.~G.
\newblock {Picture Coding Using Pseudo-Random Noise}.
\newblock \emph{IRE Transactions on Information Theory}, 1962.

\bibitem[Rényi(1953)]{renyi1953os}
Rényi, A.
\newblock On the theory of order statistics.
\newblock \emph{Acta Mathematica Hungarica}, 4:\penalty0 191--–231, 1953.

\bibitem[Schuchman(1964)]{schuchman1964dither}
Schuchman, L.
\newblock Dither signals and their effect on quantization noise.
\newblock \emph{IEEE Transactions on Communication Technology}, 12\penalty0
  (4):\penalty0 162--165, 1964.

\bibitem[Shah et~al.(2022)Shah, Chen, Balle, Kairouz, and Theis]{shah2021dp}
Shah, A., Chen, W.-N., Balle, J., Kairouz, P., and Theis, L.
\newblock Optimal compression of locally differentially private mechanisms.
\newblock In \emph{Artificial Intelligence and Statistics}, 2022.
\newblock URL \url{https://arxiv.org/abs/2111.00092}.

\bibitem[Song et~al.(2016)Song, Cuff, and Poor]{song2016ld}
Song, E.~C., Cuff, P., and Poor, H.~V.
\newblock The likelihood encoder for lossy compression.
\newblock \emph{IEEE Transactions on Information Theory}, 62\penalty0
  (4):\penalty0 1836--1849, 2016.
\newblock \doi{10.1109/TIT.2016.2529657}.

\bibitem[Sriperumbudur et~al.(2009)Sriperumbudur, Fukumizu, Gretton,
  Schölkopf, and Lanckriet]{sriperumbudur2009integral}
Sriperumbudur, B.~K., Fukumizu, K., Gretton, A., Schölkopf, B., and Lanckriet,
  G. R.~G.
\newblock On integral probability metrics, $\phi$-divergences and binary
  classification, 2009.

\bibitem[Theis \& Agustsson(2021)Theis and Agustsson]{theis2021stochastic}
Theis, L. and Agustsson, E.
\newblock On the advantages of stochastic encoders.
\newblock In \emph{Neural Compression Workshop at ICLR 2021}, 2021.

\bibitem[Townsend et~al.(2019)Townsend, Bird, and Barber]{townsend2019bb}
Townsend, J., Bird, T., and Barber, D.
\newblock Practical lossless compression with latent variables using bits back
  coding.
\newblock \emph{arXiv preprint arXiv:1901.04866}, 2019.

\bibitem[Wallace(1990)]{wallace1990bb}
Wallace, C.~S.
\newblock Classification by minimum-message-length inference.
\newblock In \emph{International Conference on Computing and Information}, pp.\
   72--81. Springer, 1990.

\bibitem[Wyner(1975)]{wyner1975ci}
Wyner, A.
\newblock The common information of two dependent random variables.
\newblock \emph{IEEE Transactions on Information Theory}, 21\penalty0
  (2):\penalty0 163--179, 1975.
\newblock \doi{10.1109/TIT.1975.1055346}.

\bibitem[Xu et~al.(2011)Xu, Liu, and Chen]{xu2011wyner}
Xu, G., Liu, W., and Chen, B.
\newblock Wyners common information for continuous random variables - a lossy
  source coding interpretation.
\newblock In \emph{45th Annual Conference on Information Sciences and Systems},
  pp.\  1--6, 2011.
\newblock \doi{10.1109/CISS.2011.5766249}.

\bibitem[Zamir(2014)]{zamir2014book}
Zamir, R.
\newblock \emph{{Lattice Coding for Signals and Networks}}.
\newblock Cambridge University Press, 2014.

\bibitem[Zamir \& Feder(1992)Zamir and Feder]{zamir1992universal}
Zamir, R. and Feder, M.
\newblock On universal quantization by randomized uniform/lattice quantizers.
\newblock \emph{IEEE Transactions on Information Theory}, 38\penalty0
  (2):\penalty0 428--436, 1992.

\bibitem[Ziv(1985)]{ziv1985universal}
Ziv, J.
\newblock On universal quantization.
\newblock \emph{IEEE Transactions on Information Theory}, 31\penalty0
  (3):\penalty0 344--347, 1985.

\end{thebibliography}

\newpage
\appendix
\onecolumn

\section{Pseudocode for greedy rejection sampling}

\begin{algorithm}[h!]
  \caption{Greedy rejection sampling \citep[RS*;][]{harsha2007}}
  \label{alg:harsha}
  \begin{algorithmic}[1]
  \Require $\texttt{p}, \texttt{q}$
    \State $n \gets 0$
    \State $\texttt{q'} \gets \texttt{0}$
    \Comment{Portion of distribution simulated thus far}
    \Statex
    \Repeat
      \State $n \gets n + 1$
      \State $z \gets \texttt{simulate}(n, \texttt{p})$
      \Comment{Generate candidate}
      \State $\texttt{p'} \gets (1 - \texttt{sum}(\texttt{q'})) \cdot \texttt{p}$ 
      \Comment{Scaled proposal distribution}
      \State $\texttt{a} \gets \texttt{min}(1, (\texttt{q} - \texttt{q'}) / \texttt{p'})$
      \Comment{Acceptance probability}
      \State $\texttt{q'} \gets \texttt{q'} + \texttt{a} \cdot \texttt{p'}$
    \Until{$\texttt{uniform}(n) < \texttt{a}(z)$}
    \Comment{Accept/reject candidate}
    \Statex
    \State \textbf{return} $n$
  \end{algorithmic}
\end{algorithm}
Algorithm~\ref{alg:harsha} (RS*) contains pseudocode for the approach taken by \citet{harsha2007} to prove a one-shot achievability result for the coding cost of reverse channel coding, the ``one-shot reverse Shannon theorem''. Our algorithm computes slightly different intermediate results and also differs in notation from the proofs of \citet{harsha2007}. For pseudocode closer in style to the paper, see Appendix~A of \citet{havasi2018miracle}.

\begin{wrapfigure}{r}{6cm}
    \centering
        \begin{tikzpicture}
        \begin{axis}[
                width=8cm,
                xlabel={$z$},
                xmin=0,
                xmax=1,
                ymin=0,
                ymax=1,
                axis lines=left,
                xtick=\empty,
                ytick=\empty,
            ]
            \addplot graphics[xmin=0,xmax=1,ymin=0,ymax=1] {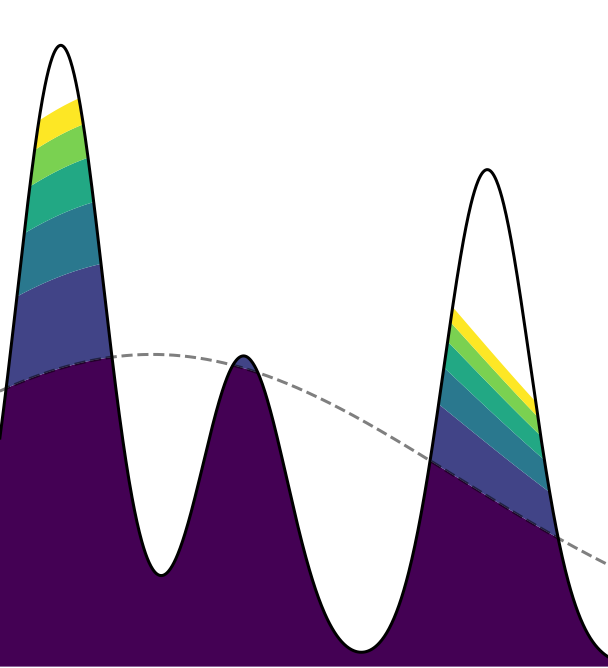};
        \end{axis}
    \end{tikzpicture}
    \vspace{-0.5cm}
    \caption{Visualization of 6 iterations of RS*. The solid line corresponds to the target distribution \texttt{q} while the dashed line indicates the proposal distribution \texttt{p}. Shaded regions correspond to $\texttt{a} \cdot \texttt{p'}$.}
    \label{fig:harsha}
\end{wrapfigure}

The algorithm effectively divides a distribution into slices and each iteration attempts to sample from one of these slices. The success probability in each iteration is proportional to the probability mass contained in a slice.

For an intuitive understanding of the algorithm, assume that \texttt{p} and \texttt{q} are vectors representing categorical proposal and target distributions, respectively. At the beginning of an iteration, \texttt{q'} represents the portion of the target distribution which we have already attempted to sample from (a sum of previously considered slices). We always have $\texttt{q'}(z) \leq \texttt{q}(z)$ and $\texttt{sum}(\texttt{q'})$ is the probability of having accepted a candidate by the current iteration. $\texttt{p'}(z)$ is thus the probability of reaching the $n$th iteration \textit{and} producing the candidate $z$. The vector $\texttt{a}$ provides an acceptance probability for each value of the candidate so that $\texttt{a} \cdot \texttt{p'}$ gives the probability of reaching the $n$th iteration, sampling a candidate value, and accepting it.

The main difference between the algorithm described above and rejection sampling is in the calculation of the acceptance probability. If we used the acceptance probability
\begin{align}
      \texttt{a} \gets \texttt{min}(1, w_\text{min} \cdot (\texttt{q} - \texttt{q'}) / \texttt{p'})
\end{align}
instead, the algorithm would already reduce to rejection sampling. Note that in this case \texttt{min} would not be needed since $w_\text{min}$ already makes sure that the acceptance probabilities do not exceed 1. Not using $w_\text{min}$ allows RS* to accept candidates with higher probability. The algorithm corrects for this ``greedy'' approach by targeting $\texttt{q} - \texttt{q'}$ instead of $\texttt{q}$ in each iteration. The algorithm is visualized in Figure~\ref{fig:harsha}.

Note that RS* requires integration and pointwise multiplication of vectors or functions where other algorithms only require evaluation of densities at a single point. In practice, this means RS* is computationally more demanding and more difficult to implement, especially for continuous distributions.

\newpage

\section{Pseudocode for ordered random coding}

\begin{algorithm}[h!]
  \caption{Ordered random coding (ORC)}
  \label{alg:orc}
  \begin{algorithmic}[1]
  \Require $\texttt{p}, \texttt{q}, w_\text{min}, N$
    \State $t, n, s^* \gets 0, 1, \infty$
    \Statex
    \Repeat
      \State $z \gets \texttt{simulate}(n, \texttt{p})$
      \Comment{Candidate generation}
      \Statex
      \State $v \gets N / (N - n + 1)$
      \State $t \gets t + v \cdot \texttt{expon}(n, 1)$
      \Comment{Candidate scoring}
      \State $s \gets t \cdot \texttt{p}(z) / \texttt{q}(z)$
      \Statex
      \If{$s < s^*$}
        \Comment{Accept/reject candidate}
        \State $s^* \gets s$
        \State $n^* \gets n$
      \EndIf
      \Statex
      \State $n \gets n + 1$
    \Until{$s^* \leq t \cdot w_\text{min}$ \textbf{or} $n > N$}
    \Statex
    \State \textbf{return} $n^*$
  \end{algorithmic}
\end{algorithm}

Algorithm~\ref{alg:orc} contains pseudocode for ordered random coding. In contrast to the Poisson functional representation, the algorithm considers only a finite number of candidates $N$, which means that it can still be used in practice when $\wmin = 0$ (either because the density ratio is unbounded or a bound is unknown). Another difference is that the exponential variables are weighted.

We note that for better numerical accuracy, log-density ratios and log-sum-exp operations should be used in practice where the pseudocode uses density ratios and sums (lines 5 and 6).

\newpage
\section{Proof of Theorem~\ref{th:orc}}

\setcounter{recounter}{3}
\setcounter{retheorem}{0}
\begin{retheorem}
Let $\Z_n \sim p$ and $G_n \sim \textnormal{Gumbel}(0, 1)$ for $n \in \{1, \dots, N\}$ (i.i.d.). Let $\tilde G_n$ be the result of sorting the random variables in decreasing order such that $\tilde G_1 \geq \dots \geq \tilde G_N$. We further define
\begin{align}
    \idx &= \argmax_{n \leq N}\log \frac{q_\x(\Z_n)}{p(\Z_n)} + G_n, 
    &
    \tildeidx &= \argmax_{n \leq N}\log \frac{q_\x(\Z_n)}{p(\Z_n)} + \tilde G_n.
\end{align}
Then $\Z_{\idx} \sim \Z_{\tildeidx}$.
\end{retheorem}
\begin{proof}
    First, let us define the functions
    \begin{align}
        n^*(\z_1, \dots, \z_N, g_1, \dots, g_N) = \text{argmax}_n \log\frac{q_\x(\z_n)}{p(\z_n)} + g_n
    \end{align}
    and
    \begin{align}
        z(\z_1, \dots, \z_N, g_1, \dots, g_N) = \z_{n^*(\z_1, \dots, \z_N, g_1, \dots, g_N)}.
    \end{align}
    It is not difficult to see that
    \begin{align}
      z(\z_{\sigma(1)}, \dots, \z_{\sigma(N)}, g_{\sigma(1)}, \dots, g_{\sigma(N)})
      &= z(\z_1, \dots, \z_N, g_1, \dots, g_{N})
    \end{align}
    for any permutation $\sigma$ of the indices $1, \dots, N$. Also note that since the candidates are i.i.d. and therefore exchangeable, permuting the candidates does not change their distribution,
    \begin{align}
        \Z_1, \dots, \Z_N \sim \Z_{\sigma(1)}, \dots, \Z_{\sigma(N)} \sim \prod_n p,
    \end{align}
    and that this is true for any permutation that is independent of $\Z_1, \dots, \Z_n$ even if the permutation depends on the $G_n$. We therefore have
    \begin{align}
      z(\Z_{\sigma(1)}, \dots, \Z_{\sigma(N)}, G_1, \dots, G_N) \sim z(\Z_1, \dots, \Z_N, G_1, \dots, G_N).
    \end{align}
    
    Choose $\tilde\sigma$ such that $\tilde G_n = G_{\tilde\sigma(n)}$. Then
    \begin{align}
      \Z_{\tildeidx}
      &= z(\Z_1, \dots, \Z_N, G_{\tilde\sigma(1)}, \dots, G_{\tilde\sigma(N)}) \\
      &= z(\Z_{\tilde\sigma^{-1}(1)}, \dots, \Z_{\tilde\sigma^{-1}(N)}, G_1, \dots, G_N) \\
      &\sim z(\Z_1, \dots, \Z_N, G_1, \dots, G_N) \\
      &= \Z_\idx.
    \end{align}
\end{proof}

\section{Proof of Corollary~\ref{th:orc_tvd}}

We first prove the following result which follows from the results of \citet{havasi2018miracle} and \citet{chatterjee2018is} and does not assume bounded density ratios.
\vspace{6pt}
\begin{lemma}
    Let $\tilde q_\x$ be the distribution of $\Z_{\tildeidx}$ where $\tildeidx$ is defined as in Theorem~\ref{th:orc}. If the number of candidates is $N = 2^{\KL{q_\x}{p} + t}$ for some $t \geq 0$ and $\Z \sim q_\x$, then
    \begin{align}
        \TV{\tilde q}{q} \leq 4\epsilon
        \label{eq:tvd_4eps}
    \end{align}
    where 
    \begin{align}
        \epsilon = \left( 2^{-t / 4} + 2 \sqrt{\mathbb{P}\left(\log \frac{q_\x(\Z)}{p(\Z)} > \KL{q_\x}{p} + t / 2\right)} \right)^\frac{1}{2}.
    \end{align}
    \label{th:tv_lemma}
\end{lemma}
\begin{proof}
    If $\varepsilon \geq 1/4$ then Eq.~\ref{eq:tvd_4eps} is automatically true and there is nothing left to show. Assume therefore that $q, p,$ and $t$ are such that $\varepsilon < 1/4$.
    
    Since $\Z_\tildeidx \sim \Z_\idx$ by Theorem~\ref{th:orc}, $\tilde q_\x$ is also the distribution of $\Z_\idx$.
    Let $\Omega = \{\Z_1, \dots, \Z_N\}$ be the set of candidates and let $\tilde q_{\x,\Omega}$ be the distribution of $\Z_\idx$ for a fixed set of candidates. That is,
    \begin{align}
        \tilde q_{\x,\Omega}(\z) = \sum_{n = 1}^N \pi(n) \delta(\z - \Z_n)
    \end{align}
    where $\pi(n) \propto q_\x(\Z_n) / p(\Z_n)$.
    Theorem 3.2 of \citet{havasi2018miracle} tells us that
    \begin{align}
        \mathbb{P}\left(\left|\mathbb{E}_{q_{\x,\Omega}}[f(\tilde\Z)] - \mathbb{E}_{q_\x}[f(\Z)]\right| \geq 2 \|f\|_q \frac{\epsilon}{1 - \epsilon}\right) < 2\epsilon,
        \label{eq:th32}
    \end{align}
    for any measurable function $f$, where $\|f\|_q = \sqrt{\mathbb{E}_{q_\x}[f(\Z)^2]}$ and the probability arises due to the randomness in the set of candidates $\Omega$. We choose
    \begin{align}
       f(\z) = \begin{cases}
        1 & \text{ if } \tilde q_\x(\z) > q_\x(\z), \\
        -1 & \text{ else. }
       \end{cases} 
    \end{align}
    For notational convenience, we further define the event
    \begin{align}
        A = \left[\!\!\left[ \left|\mathbb{E}_{q_{\x,\Omega}}[f(\tilde\Z)] - \mathbb{E}_{q_\x}[f(\Z)]\right| \geq \|f\|_q \frac{2\epsilon}{1 - \epsilon} \right]\!\!\right],
    \end{align}
    where $[\![ \cdot ]\!]$ is 1 if its argument is true and 0 otherwise. We have
    \begin{align}
        2\TV{\tilde q}{q}
        &= |\mathbb{E}_{\tilde q}[f(\tilde\Z)] - \mathbb{E}_{q_\x}[f(\Z)]| \label{eq:tvd0} \\
        &= |\mathbb{E}_\Omega[\mathbb{E}_{\tilde q_{\x,\Omega}}[f(\tilde\Z)]] - \mathbb{E}_{q_\x}[f(\Z)]| \\
        &\leq \mathbb{E}_\Omega[|\mathbb{E}_{\tilde q_{\x,\Omega}}[f(\tilde\Z)] - \mathbb{E}_{q_\x}[f(\Z)]|]  \label{eq:tvd1} \\
        &= P(A = 1)\mathbb{E}_\Omega[|\mathbb{E}_{\tilde q_{\x,\Omega}}[f(\tilde\Z)] - \mathbb{E}_{q_\x}[f(\Z)]| \mid A = 1]
        + P(A = 0)\mathbb{E}_\Omega[|\mathbb{E}_{\tilde q_{\x,\Omega}}[f(\tilde\Z)] - \mathbb{E}_{q_\x}[f(\Z)]| \mid A = 0] \\
        &\leq \textstyle 2 P(A = 1) + (1 - P(A = 1)) \frac{2\epsilon}{1 - \epsilon} \|f\|_q  \label{eq:tvd2} \\
        &= \textstyle 2 P(A = 1) \left( 1 - \frac{\epsilon}{1 - \epsilon} \right) + \frac{2\epsilon}{1 - \epsilon} \label{eq:tvd3}  \\
        &\leq \textstyle 4\epsilon  \left( 1 - \frac{\epsilon}{1 - \epsilon} \right) + \frac{2\epsilon}{1 - \epsilon} \label{eq:tvd4} \\
        &\leq 4\epsilon + 4 \epsilon \\
        &= 8\epsilon,
    \end{align}
    where Eq.~\ref{eq:tvd0} is a known identity \citep{sriperumbudur2009integral}, Eq.~\ref{eq:tvd1} follows from Jensen's inequality, Eq.~\ref{eq:tvd2} follows from the definitions of $f$ and $A$, 
    Eq.~\ref{eq:tvd3} follows from \hbox{$\|f(\z)\|_q = 1$}, and
    Eq.~\ref{eq:tvd4} follows from Eq.~\ref{eq:th32}.
\end{proof}
\vspace{6pt}
\setcounter{recounter}{3}
\setcounter{recorollary}{1}
\begin{recorollary}
    Let $\tilde q_\x$ be the distribution of $\Z_{\tildeidx}$ where $\tildeidx$ is defined as in Theorem~\ref{th:orc}. If the number of candidates is $N = 2^{\KL{q_\x}{p} + t}$ for some $t \geq 2 e^{-1}\log e$ and $p(\z) / q_\x(\z) \geq \wmin > 0$ for all $\z$, then
    \begin{align}
        \TV{\tilde q_\x}{q_\x} \leq 4\varepsilon
    \end{align}
    where with $B = -\log \wmin$ we have
    \begin{align}
         \epsilon \leq 2^{-t/8} + \sqrt{2} \exp\left( -\frac{1}{4B^2}(t/2 - e^{-1} \log e)^2 \right).
    \end{align}
\end{recorollary}
\begin{proof}
    By Lemma~\ref{th:tv_lemma}, we have
    \begin{align}
        \TV{\tilde q_\x}{q_\x} \leq 4\epsilon
    \end{align}
    where 
    \begin{align}
        \epsilon = \left( 2^{-t / 4} + 2 \sqrt{\mathbb{P}\left(\log \frac{q_\x(\Z)}{p(\Z)} > \KL{q_\x}{p} + t / 2\right)} \right)^\frac{1}{2}.
    \end{align}
    To prove our claim we need to bound $\epsilon$. Define
    \begin{align}
        l(\z) = \max(0, \log q_\x(\z) - \log p(\z)).
    \end{align}
    By Claim A.2 of \citet{harsha2007}, we have
    \begin{align}
        \mathbb{E}_{q_\x}[l(\Z)]
        = \mathbb{E}_{q_\x}\left[ \max\left(0, \log \frac{q_\x(\Z)}{p(\Z)}\right) \right]
        = \mathbb{E}_{q_\x}\left[ \log \frac{q_\x(\Z)}{p(\Z)} - \min\left(0, \log \frac{q_\x(\Z)}{p(\Z)}\right) \right]
        \leq \KL{q_\x}{p} + e^{-1}\log e.
        \label{eq:claima2}
    \end{align}
    Let $B = -\log \wmin$ so that $l(\z) \leq B$ for all $\z$. We have
    \begin{align}
        \mathbb{P}\left(\log \frac{q_\x(\Z)}{p(\Z)} > \KL{q_\x}{p} + t / 2\right)
        &= \mathbb{P}\left(l(\Z) > \KL{q_\x}{p} + t / 2\right) \\
        &= \mathbb{P}\left(l(\Z) - \mathbb{E}_{q_\x}[l(\Z)] > \KL{q_\x}{p} - \mathbb{E}_{q_\x}[l(\Z)] + t / 2\right) \\
        &\leq \exp\left( -(\KL{q_\x}{p} - \mathbb{E}_{q_\x}[l(\Z)] + t/2)^2 / B^2 \right) \label{eq:hoeff} \\
        &\leq \exp\left( -(t/2 - e^{-1} \log e)^2 / B^2 \right) \label{eq:epsbound0}
    \end{align}
    where the first equality follows from the non-negativity of the KL divergence, Eq.~\ref{eq:hoeff} follows from Hoeffding's inequality, and the last inequality follows from Eq.~\ref{eq:claima2} and $t/2 \geq e^{-1}\log e$. Thus, we have
    \begin{align}
        \epsilon 
        &\leq \left(2^{-t/4} + 2 \exp\left( -\frac{1}{2B^2}(t/2 - e^{-1} \log e)^2 \right)\right)^{\frac{1}{2}} \\
        &\leq 2^{-t/8} + \sqrt{2} \exp\left( -\frac{1}{4B^2}(t/2 - e^{-1} \log e)^2 \right)
    \end{align}
    where the second inequality follows from $\sqrt{a + b} \leq \sqrt{a} + \sqrt{b}$.
\end{proof}

\section{Proof of Theorem~\ref{th:orc_pfr}}

\setcounter{recounter}{3}
\setcounter{retheorem}{2}
\begin{retheorem}
    Let $S_n$ be exponentially distributed RVs and $\Z_n \sim p$ for all $n \in \mathbb{N}$ (i.i.d.), and let 
    \begin{align}
        T_n &= \sum_{m = 1}^n S_m, &
        \tilde T_{N,n} &= \sum_{m = 1}^n \frac{N}{N - m + 1} S_m
    \end{align}
    for $N \in \mathbb{N}$. Let
    \begin{align}
        \idxpfr &= \argmin_{n \in \mathbb{N}} T_n \frac{p(\Z_n)}{q_\x(\Z_n)}, \\
        \idxsis &= \argmin_{n \leq N} \tilde T_{N,n} \frac{p(\Z_n)}{q_\x(\Z_n)}.
    \end{align}
    Then $\idxsis \leq \idxpfr$. Further, if $\idxpfr$ is finite then there exists an $M \in \mathbb{N}$ such that for all $N \geq M$ we have $\idxsis = \idxpfr$.
\end{retheorem}
\begin{proof}
    We first show that $\idxsis \leq \idxpfr$. For any $\Delta \geq 0$ and $n$ with $n + \Delta \leq N$ we have
    \begin{align}
        \frac{\tilde T_{N,n + \Delta}}{\tilde T_{N,n}}
        \label{eq:rate1}
        &= \frac{\sum_{m = 1}^{n + \Delta} \frac{N}{N - m + 1} S_m}{\sum_{m = 1}^n \frac{N}{N - m + 1} S_m} \\
        &= \frac{\sum_{m = 1}^{n + \Delta} \frac{N - n +1}{N - m + 1} S_m}{\sum_{m = 1}^n \frac{N - n + 1}{N - m + 1} S_m} \\
        &= \frac{\sum_{m = 1}^{n} \frac{N - n + 1}{N - m + 1} S_m + \sum_{m = n + 1}^{n + \Delta} \frac{N - n + 1}{N - m + 1} S_m}{\sum_{m = 1}^n \frac{N - n + 1}{N - m + 1} S_m} \\
        &\geq \frac{\sum_{m = 1}^{n} \frac{N - n + 1}{N - m + 1} S_m + \sum_{m = n + 1}^{n + \Delta} S_m}{\sum_{m = 1}^n \frac{N - n + 1}{N - m + 1} S_m} \label{eq:inequ1} \\
        &\geq \frac{\sum_{m = 1}^{n} S_m + \sum_{m = n + 1}^{n + \Delta} S_m}{\sum_{m = 1}^n S_m} \label{eq:inequ2} \\
        &= \frac{T_{n + \Delta}}{T_n}
        \label{eq:rate2}
    \end{align}
    That is, $\tilde T_{N,n}$ increases at a faster rate than $T_n$. Eq.~\ref{eq:inequ1} is true because $S_m \geq 0$ and $(N - n + 1)/(N - m + 1) > 1$ for $m \geq n + 1$. Eq.~\ref{eq:inequ2} is true because $(N - n + 1)/(N - m + 1) \leq 1$ for $m \leq n$ and 
    \begin{align}
        f: [0, \infty) \rightarrow [0, \infty), \quad t \mapsto \frac{t + a}{t}
    \end{align}
    is a monotonically decreasing function for any $a \geq 0$.
    
    Now assume that $\idxpfr$ takes on some value $\nstar \in \mathbb{N}$. If $\nstar \geq N$ then $\idxsis \leq \idxpfr$ since $\idxsis \leq N$. If $\nstar < N$, to show that $\idxsis$ does not exceed $\idxpfr$ it is enough to show that\footnote{We assume that in case of a tie, argmin returns the smaller index.}
    \begin{align}
        \tilde T_{N,\nstar} \frac{p(\Z_\nstar)}{q_\x(\Z_\nstar)}
        \leq \tilde T_{N,\nstar + \Delta} \frac{p(\Z_{\nstar + \Delta})}{q_\x(\Z_{\nstar + \Delta})}
        \label{eq:sis_scores}
    \end{align}
    for any $\Delta > 0 $ with $\nstar + \Delta \leq N$, since $\idxsis$ then must be either $\nstar$ or take on a smaller value.
    By definition of $\idxpfr$, we have
    \begin{align}
        T_\nstar \frac{p(\Z_\nstar)}{q_\x(\Z_\nstar)} &\leq T_{\nstar + \Delta} \frac{p(\Z_{\nstar + \Delta})}{q_\x(\Z_{\nstar + \Delta})}
        \label{eq:premise}
    \end{align}
    for any $\Delta > 0$. From this and Eqs.~\ref{eq:rate1} to \ref{eq:rate2} it follows that
    \begin{align}
        \frac{p(\Z_\nstar)}{q_\x(\Z_\nstar)} &\leq \frac{T_{\nstar + \Delta}}{T_\nstar} \frac{p(\Z_{\nstar + \Delta})}{q_\x(\Z_{\nstar + \Delta})}, \\
        \frac{p(\Z_\nstar)}{q_\x(\Z_\nstar)} &\leq \frac{\tilde T_{N,\nstar + \Delta}}{\tilde T_{N,\nstar}} \frac{p(\Z_{\nstar + \Delta})}{q_\x(\Z_{\nstar + \Delta})}, \\
        \tilde T_{N,n} \frac{p(\Z_\nstar)}{q_\x(\Z_\nstar)} &\leq \tilde T_{N,{\nstar + \Delta}} \frac{p(\Z_{\nstar + \Delta})}{q_\x(\Z_{\nstar + \Delta})}, \\
    \end{align}
    which concludes the proof of $\idxsis \leq \idxpfr$.
    
    Next, we show that $\idxsis = \idxpfr$ for large enough $N$. First note that we can equivalently define $\idxsis$ as follows,
    \begin{align}
        N' &= \min \{ \idxpfr, N \}, &
        \idxsis &= \argmin_{n \leq N'} \tilde T_{N,n} \frac{p(\Z_n)}{q_\x(\Z_n)}.
    \end{align}
    We have
    \begin{align}
        \lim_{N \rightarrow \infty} \tilde T_{N,n}
        = \sum_{m = 1}^n \left( \lim_{N \rightarrow \infty} \frac{N}{N - m + 1} \right) S_m 
        = T_n
    \end{align}
    for all $n \leq \idxpfr$ and therefore 
    \begin{align}
        \lim_{N \rightarrow \infty} \idxsis
        &= \lim_{N \rightarrow \infty} \argmin_{n \leq N'} \tilde T_{N,n} \frac{p(\Z_n)}{q_\x(\Z_n)} \\
        &= \argmin_{n \leq N'} \lim_{N \rightarrow \infty} \tilde T_{N,n} \frac{p(\Z_n)}{q_\x(\Z_n)} \\
        &= \argmin_{n \leq \idxpfr} \lim_{N \rightarrow \infty} \tilde T_{N,n} \frac{p(\Z_n)}{q_\x(\Z_n)} \\
        &= \argmin_{n \leq \idxpfr} T_n \frac{p(\Z_n)}{q_\x(\Z_n)} \\
        &= \idxpfr.
    \end{align}
    Hence, since $\idxpfr$ is assumed to be finite, there is an $M \in \mathbb{N}$ such that $\idxsis = \idxpfr$ for $N \geq M$.
\end{proof}

\section{Proof of Corollary~\ref{th:orc_entropy}}

\setcounter{recounter}{3}
\setcounter{recorollary}{3}
\begin{recorollary}
    Let $C = \mathbb{E}_\X[\KL{q_\X}{p}]$ and let $\idxsis$ be defined as in Theorem~\ref{th:orc_pfr}. Then
    \begin{align}
        H[\idxsis] < C + \log (C + 1) + 4.
    \end{align}
\end{recorollary}
\begin{proof}
    Let $\idxpfr$ be defined as in Theorem~\ref{th:orc_pfr}. \citet[][Appendix A]{li2018pfr} showed that
    \begin{align}
        \mathbb{E}[\log \idxpfr \mid \X = \x] \leq \KL{q_\x}{p} + e^{-1} \log e + 1.
    \end{align}
    While \citet{li2018pfr} were only considering the case where $p(\z) = \mathbb{E}_\X[q_\X(\z)]$, their proof of the above statement does not make use of this assumption. Since $\idxsis \leq \idxpfr$, we also have
    \begin{align}
        \mathbb{E}[\log \idxsis] \leq \mathbb{E}[\log \idxpfr] \leq \mathbb{E}_\X[\KL{q_\X}{p}] + e^{-1} \log e + 1.
        \label{eq:sis_log_bound}
    \end{align}
    \citet[][Appendix B]{li2018pfr} further showed that for any random variable $\idx$ with values in $\mathbb{N}$, we have
    \begin{align}
        \mathbb{E}[-\log p_\lambda(\idx)]
        \leq \mathbb{E}[\log \idx] + \log (\mathbb{E}[\log \idx] + 1) + 1,
    \end{align}
    where $p_\lambda(n) \propto n^{-\lambda}$ is a Zipf distribution with
    \begin{align}
        \lambda = 1 + 1/\EX[\log \idx].
    \end{align}
    Applied to $\idxsis$ and using Eq.~\ref{eq:sis_log_bound}, we get
    \begin{align}
        H[\idxsis]
        &\leq \mathbb{E}[-\log p_\lambda(\idxsis)] \\
        &\leq \mathbb{E}[\log \idxsis] + \log (\mathbb{E}[\log \idxsis] + 1) + 1 \\
        &\leq C + \log (C + e^{-1} \log e + 2) + e^{-1} \log e + 2 \\
        &< C + \log (C + 1) + 4.
    \end{align}
\end{proof}

\section{Proof of Theorem~\ref{th:hybrid}}

Let us first briefly repeat the relevant definitions from the main text. We have
\begin{align}
    r_\x(\z) = \begin{cases}
        1 & \text{ if } \z \in \cx + [-0.5, 0.5)^D, \\
        0 & \text{ else, }
    \end{cases}
\end{align}
where $\cx$ is chosen such that the support of $r_\x$ is contained within $[0, M_1) \times \cdots [0, M_D)$.
Candidates are generated via dithered quantization
\begin{align}
    \mathbf{U}_n &\sim \text{Uniform}([0, 1)^D), &
    \mathbf{K}_n &= \lfloor \cx - \mathbf{U}_n \rceil, &
    \mathbf{Z}_n &= \mathbf{K}_n + \mathbf{U}_n,
\end{align}
so that $\mathbf{Z}_n \sim r_\x$. One of the candidates is then selected according to
\begin{align}
    \tilde T_{N,n} &= \sum_{m = 1}^n \frac{N}{N - m + 1} S_m, &
    \idx &= \argmin_{n \leq N} \tilde T_{N,n} \frac{r_\x(\Z_n)}{q_\x(\Z_n)},
    \label{eq:idx_hybrid2}
\end{align}
where the support of $q_\x$ is assumed to be contained in the support of $r_\x$. For notational convenience, further define 
\begin{align}
    \bks &= \mathbf{K}_\idx, &
    \label{eq:bks_hybrid2}
    \mathbf{U}^* &= \mathbf{U}_\idx, &
    \mathbf{Z}^* &= \mathbf{Z}_\idx.
\end{align}
Note that $\mathbf{Z}^* = \bks + \mathbf{U}^*$. The following theorem bounds the coding cost of optimally encoding $\idx$ and $\bks$.
\vspace{8pt}
\setcounter{recounter}{3}
\setcounter{retheorem}{4}
\begin{retheorem}
    Let $\idx$ and $\bks$ be defined as in Eqs.~\ref{eq:idx_hybrid2} and \ref{eq:bks_hybrid2} and let $p$ be the uniform distribution over $[0, M_1) \times \cdots \times [0, M_D)$ for some $M_i \in \mathbb{N}$. Then
    \begin{align*}
        H[\idx, \bks]
        < C + \log (C - \textstyle\sum_i \log M_i + 1) + 4,
    \end{align*}
    where $C = \mathbb{E}_\X[\KL{q_\X}{p}]$.
\end{retheorem}
\begin{proof}
    By Corollary~\ref{th:orc_entropy}, we have
    \begin{align}
        H[\idx] < C' + \log(C' + 1) + 4
    \end{align}
    where $C' = \mathbb{E}_\X[\KL{q_\X}{r_\X}] = C - \sum_i \log M_i$, assuming the support of $q_\x$ is contained in the support of $r_\x$.
    
    Next consider the coding cost of $\bks$.
    Note that for each entry $K_i^*$ in $\bks$ we must have $0 \leq K_i^* < M_i$ since otherwise $K_i^* + U_i^* < 0$ or $K_i^* + U_i^* \geq M_i$, that is, $\Z^* = \bks + \mathbf{U}^*$ would be outside the support of $r_\X$. Hence,
    \begin{align}
        H[\bks] 
        &\leq \sum_i \log M_i  \\
        &= \log \frac{1}{p(\z)} \\
        &= \mathbb{E}_\X\left[\mathbb{E}_{\Z \sim q_\X}\left[\log \frac{r_\X(\Z)}{p(\Z)}\right]\right] \\
        &= \mathbb{E}_\X\left[\mathbb{E}_{\Z \sim q_\X}\left[\log \frac{q_\X(\Z)r_\X(\Z)}{p(\Z)q_\X(\Z)}\right]\right] \\
        &= C - C'.
    \end{align}
    Taken together, we have
    \begin{align}
        H[\idx, \bks] \leq H[\idx] + H[\bks] < C' + \log(C' + 1) + 4 + C - C' = C + \log(C - \sum_i \log M_i + 1) + 4,
    \end{align}
    proving the claim.
\end{proof}

\section{Hybrid coding for Gaussian distributions}

Let $\tilde q_\x$ be a truncated Gaussian with mean $\x$ and covariance $\mathbf{I}$ and let $\theta$ be the fraction of mass which has been truncated. We assume that the mean is itself Gaussian distributed with covariance $\sigma^2 \mathbf{I}$ so that
a Gaussian with covariance $(\sigma^2 + 1)\mathbf{I}$ is a suitable candidate generating distribution $\tilde p$.
 
To make the marginal distribution uniform, we first transform each coordinate with the CDF of a univariate Gaussian with variance $\sigma^2 + 1$, $\Phi_{\sigma^2 + 1}$. After this transformation, different target distributions have supports of varying widths. The distribution with the widest support is centered at zero. The support of the truncated Gaussian is limited to the left and right by
\begin{align}
    a &= \Phi^{-1}(\theta' / 2), & b &= \Phi^{-1}(1 - \theta' / 2),
\end{align}
along each coordinate, where $\theta' = 1 - (1 - \theta)^{1/D}$ and $\Phi$ is the CDF of a standard normal. After the transformation, the limits of the support become $\Phi_{\sigma^2 + 1}(a)$ and $\Phi_{\sigma^2 + 1}(b)$, respectively, so that we can scale the distributions by
\begin{align}
    M = \left\lfloor \frac{1}{\Phi_{\sigma^2 + 1}(b) - \Phi_{\sigma^2 + 1}(a)} \right\rfloor
\end{align}
along the $i$th coordinate while still ensuring that the distributions fit into a unit interval. For $D = 1$, the target distribution becomes
\begin{align}
    q_x(z)
    = \frac{\tilde q_x(\tilde z)}{M \Phi_{\sigma^2 + 1}'(\tilde z)}
    = \frac{1}{M} \frac{\tilde q_x(\tilde z)}{\mathcal{N}(\tilde z; 0, \sigma^2 + 1)}
\end{align}
where $\tilde z = \Phi_{\sigma^2 + 1}^{-1}(z / M)$ and $z \in [0, M)$.
This is illustrated in Figure~\ref{fig:hybrid}.
For $D > 1$, we have
\begin{align}
    q_\x(\z) = \prod_i q_{x_i}(z_i).
\end{align}

For $\wmin$, we choose
\begin{align}
    \inf_{\z} \frac{p(\z)}{q_\x(\z)}
    =
    \inf_{\tilde\z} \frac{\tilde p(\tilde\z)}{\tilde q_\x(\tilde\z)}
    \geq \inf_{\tilde\z} \frac{(1 - \theta) \mathcal{N}(\tilde\z; 0, (\sigma^2 + 1)\mathbf{I})}{\mathcal{N}(\tilde\z; \x, \mathbf{I})}
    = (1 - \theta)\frac{\mathcal{N}(\tilde\z_\text{min}; 0, (\sigma^2 + 1)\mathbf{I})}{\mathcal{N}(\tilde\z_\text{min}; \x, \mathbf{I})}
    = \wmin
    \label{eq:wming}
\end{align}
where
\begin{align}
    \tilde\z_\text{min} = \frac{\sigma^2 + 1}{\sigma^2} \mathbf{x}.
\end{align}
is the minimizer of the infimum on the right-hand side.
Since the density ratios are invariant under transformation, we can use the same $\wmin$ for ORC/PFR and the hybrid coding scheme. 

\newpage
\section{Additional results}
\label{sec:additional}

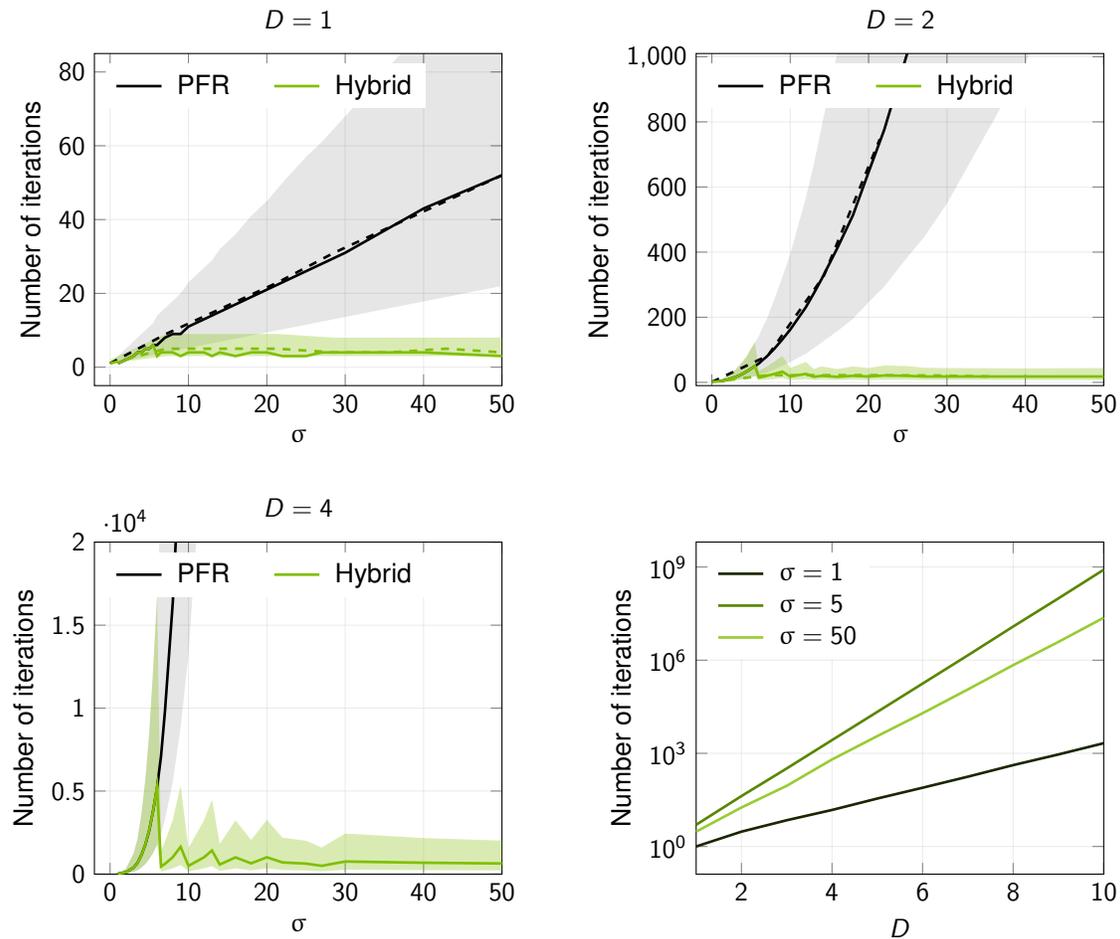
\begin{figure}[h!]
    \centering
        \begin{tikzpicture}

        \begin{axis}[
                xlabel={$\sigma$},
                ylabel={Number of iterations},
                width=7cm,
                height=6cm,
                legend columns=-1,
                legend pos=north west,
                legend style={
                    draw=none,
                    /tikz/every even column/.append style={column sep=0.5cm},
                    /tikz/every odd column/.append style={column sep=0.12cm}
                },
                ymin=-5,
                ymax=85,
                xmin=-2,
                xmax=50,
                grid,
                minor grid style={opacity=0.3},
                major grid style={opacity=0.3},
                title={$D = 1$},
            ]

            \addplot[PFR2] table[x=sigma,y=p50] {figures/data/gaussian_1_pfr.dat};
            \addlegendentry{PFR};
            \addplot[hybrid] table[x=sigma,y=p50] {figures/data/gaussian_1_hybrid.dat};
            \addlegendentry{Hybrid};
            
            \addplot[PFR2, dashed] table[x=sigma,y=p50] {figures/data/gaussian_pfr.dat};
            \addplot[hybrid, dashed] table[x=sigma,y=p50] {figures/data/gaussian_hybrid.dat};
            
            \addplot[opacity=0,name path=pfr75] table[x=sigma,y=p75] {figures/data/gaussian_1_pfr.dat};
            \addplot[opacity=0,name path=pfr25] table[x=sigma,y=p25] {figures/data/gaussian_pfr.dat};
            \addplot[PFR2,opacity=0.1] fill between[of=pfr25 and pfr75];

            \addplot[opacity=0,name path=hybrid75] table[x=sigma,y=p75] {figures/data/gaussian_hybrid.dat};
            \addplot[opacity=0,name path=hybrid25] table[x=sigma,y=p25] {figures/data/gaussian_hybrid.dat};
            \addplot[hybrid,opacity=0.3] fill between[of=hybrid25 and hybrid75];
        \end{axis}

        \begin{axis}[
                xshift=8cm,
                xlabel={$\sigma$},
                ylabel={Number of iterations},
                width=7cm,
                height=6cm,
                legend columns=-1,
                legend pos=north west,
                legend style={
                    draw=none,
                    /tikz/every even column/.append style={column sep=0.5cm},
                    /tikz/every odd column/.append style={column sep=0.12cm}
                },
                ymin=-10,
                ymax=1010,
                xmin=-2,
                xmax=50,
                grid,
                minor grid style={opacity=0.3},
                major grid style={opacity=0.3},
                title={$D = 2$},
            ]

            \addplot[PFR2] table[x=sigma,y=p50] {figures/data/gaussian_2_pfr.dat};
            \addlegendentry{PFR};
            \addplot[hybrid] table[x=sigma,y=p50] {figures/data/gaussian_2_hybrid.dat};
            \addlegendentry{Hybrid};
            
            \addplot[PFR2, dashed,select coords between index={0}{3}] table[x=sigma,y=p50] {figures/data/gaussian_pfr_2.dat};
            \addplot[hybrid, dashed] table[x=sigma,y=p50] {figures/data/gaussian_hybrid_2.dat};

            \addplot[opacity=0,name path=pfr75] table[x=sigma,y=p75] {figures/data/gaussian_2_pfr.dat};
            \addplot[opacity=0,name path=pfr25] table[x=sigma,y=p25] {figures/data/gaussian_2_pfr.dat};
            \addplot[PFR2,opacity=0.1] fill between[of=pfr25 and pfr75];

            \addplot[opacity=0,name path=hybrid75] table[x=sigma,y=p75] {figures/data/gaussian_2_hybrid.dat};
            \addplot[opacity=0,dashed,name path=hybrid25] table[x=sigma,y=p25] {figures/data/gaussian_2_hybrid.dat};
            \addplot[hybrid,opacity=0.3] fill between[of=hybrid25 and hybrid75];
        \end{axis}
        
        \begin{axis}[
                xshift=0cm,
                yshift=-6.5cm,
                xlabel={$\sigma$},
                ylabel={Number of iterations},
                width=7cm,
                height=6cm,
                legend columns=-1,
                legend pos=north west,
                legend style={
                    draw=none,
                    /tikz/every even column/.append style={column sep=0.5cm},
                    /tikz/every odd column/.append style={column sep=0.12cm}
                },
                ymin=-10,
                ymax=20010,
                xmin=-2,
                xmax=50,
                grid,
                minor grid style={opacity=0.3},
                major grid style={opacity=0.3},
                title={$D = 4$},
            ]

            \addplot[PFR2,select coords between index={0}{13}] table[x=sigma,y=p50] {figures/data/gaussian_4_pfr.dat};
            \addlegendentry{PFR};
            \addplot[hybrid] table[x=sigma,y=p50] {figures/data/gaussian_4_hybrid.dat};
            \addlegendentry{Hybrid};

            \addplot[opacity=0,name path=pfr75,select coords between index={0}{14}] table[x=sigma,y=p75] {figures/data/gaussian_4_pfr.dat};
            \addplot[opacity=0,name path=pfr25,select coords between index={0}{16}] table[x=sigma,y=p25] {figures/data/gaussian_4_pfr.dat};
            \addplot[PFR2,opacity=0.1] fill between[of=pfr25 and pfr75];

            \addplot[opacity=0,name path=hybrid75] table[x=sigma,y=p75] {figures/data/gaussian_4_hybrid.dat};
            \addplot[opacity=0,dashed,name path=hybrid25] table[x=sigma,y=p25] {figures/data/gaussian_4_hybrid.dat};
            \addplot[hybrid,opacity=0.3] fill between[of=hybrid25 and hybrid75];
        \end{axis}

        \begin{axis}[
                xshift=8cm,
                yshift=-6.5cm,
                xlabel={$D$},
                ylabel={Number of iterations},
                width=7cm,
                height=6cm,
                legend columns=1,
                legend pos=north west,
                legend cell align={left},
                legend style={
                    draw=none,
                    /tikz/every even column/.append style={column sep=0.5cm},
                    /tikz/every odd column/.append style={column sep=0.12cm}
                },
                ymin=-10,
                xmin=1,
                xmax=10,
                ymode=log,
                grid,
                minor grid style={opacity=0.3},
                major grid style={opacity=0.3},
            ]

            \addplot[color=green!50!orange!20!black] table[x=dim,y=s1] {figures/data/hybrid_dim_vs_iter.dat};
            \addlegendentry{$\sigma = 1$};

            \addplot[color=green!50!orange!70!black] table[x=dim,y=s5] {figures/data/hybrid_dim_vs_iter.dat};
            \addlegendentry{$\sigma = 5$};

            \addplot[color=green!50!orange!80!white] table[x=dim,y=s50] {figures/data/hybrid_dim_vs_iter.dat};
            \addlegendentry{$\sigma = 50$};
        \end{axis}
    \end{tikzpicture}
    \caption{Additional results comparing the computational cost of the hybrid coding scheme with PFR/ORC on multivariate Gaussian distributions (25th, median, and 75th percentile). Dashed lines correspond to empirical performance measured by running the full algorithm while solid lines were obtained through simulations by sampling the number of iterations from their known distributions.}
    \label{fig:hybrid_extra}
\end{figure}

Figure~\ref{fig:hybrid_extra} illustrates the computational cost of encoding Gaussian samples with either the Poisson functional representation (PFR) or the hybrid coding scheme ($N = \infty$). For a fixed target distribution and corresponding $\wmin$, the number of iterations of the PFR is geometrically distributed with parameter $\wmin$ \citep{maddison2016ppmc}. For the hybrid algorithm, the number of iterations is also geometrically distributed but with parameter $\wmin \prod_i M_i$. Solid lines in Figure~\ref{fig:hybrid_extra} were estimated by repeatedly sampling a target distribution $q_\x$, then computing $\wmin$ and finally sampling from the corresponding geometric distribution. Dashed lines correspond to empirical results by running the full algorithm. The empirical results match our simulated results.

For small $\sigma$ the performance of the hybrid coding scheme matches that of the PFR. Once $\sigma$ is large enough that $M_i > 1$, the hybrid coding scheme substantially outperforms the PFR. The wiggliness in the green curves is explained by the fact that the $M_i$ are discontinuous functions of $\sigma$.

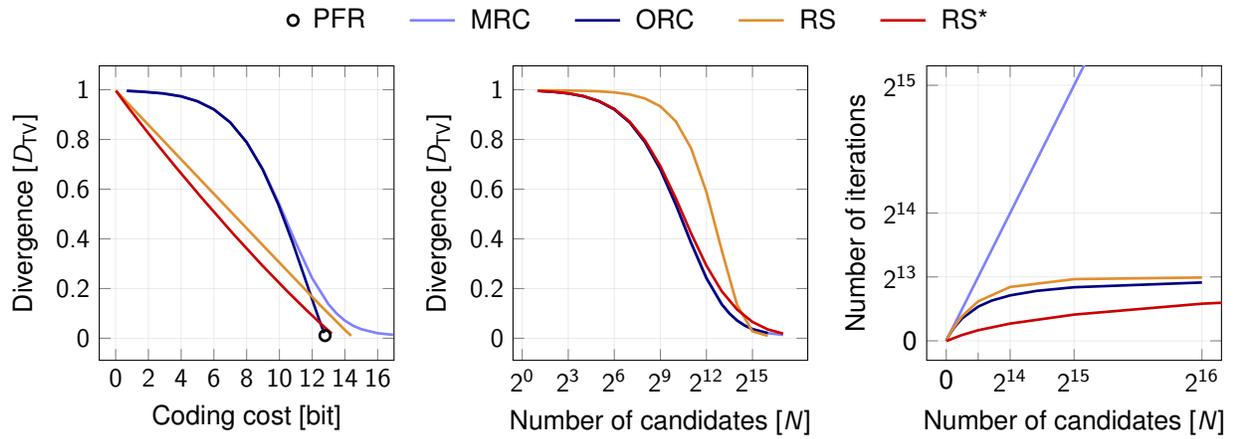
\begin{figure}[h!]
    \centering
        \begin{tikzpicture}
        \begin{axis}[
                xlabel={Coding cost [bit]},
                ylabel={Divergence [$D_\text{TV}$]},
                width=5.5cm,
                height=5.5cm,
                legend columns=-1,
                legend style={
                    at={(12cm, 4.5cm)},
                    draw=none,
                    /tikz/every even column/.append style={column sep=0.5cm},
                    /tikz/every odd column/.append style={column sep=0.12cm}
                },
                xmin=-1,
                xmax=17,
                xtick={0,2,...,16},
                ytick={0.0,0.2,0.4,0.6,0.8,1.0},
                grid,
                minor grid style={opacity=0.3},
                major grid style={opacity=0.3},
            ]
            \addplot[PFR] table[x=cc,y=TV,select coords between index={17}{17}] {figures/data/PFR.dat};
            \addlegendentry{PFR};
            
            \addplot[MRC] table[x=cc,y=TV] {figures/data/MRC.dat};
            \addlegendentry{MRC};
            
            \addplot[ORC] table[x=cc,y=TV] {figures/data/ORC.dat};
            \addlegendentry{ORC};
            
            \addplot[RS] table[x=cc,y=TV] {figures/data/RS.dat};
            \addlegendentry{RS};
            
            \addplot[Harsha] table[x=cc,y=TV] {figures/data/Harsha.dat};
            \addlegendentry{RS*};
        \end{axis}
        \begin{axis}[
                xmode=log,
                log basis x={2},
                xshift=5.5cm,
                xlabel={Number of candidates [$N$]},
                ylabel={Divergence [$D_\text{TV}$]},
                width=5.5cm,
                height=5.5cm,
                xtick={1,8,64,512,4096,32768},
                ytick={0.0,0.2,0.4,0.6,0.8,1.0},
                grid,
                major grid style={opacity=0.3},
            ]
            \addplot[MRC] table[x=N,y=TV] {figures/data/MRC.dat};
            \addplot[ORC] table[x=N,y=TV] {figures/data/ORC.dat};
            \addplot[RS] table[x=N,y=TV] {figures/data/RS.dat};
            \addplot[Harsha] table[x=N,y=TV] {figures/data/Harsha.dat};
        \end{axis}
        \begin{axis}[
                log basis x={2},
                log basis y={2},
                xshift=11cm,
                xlabel={Number of candidates [$N$]},
                ylabel={Number of iterations},
                xtick={0,16384,32768,65536},
                minor xtick={8192},
                xticklabels={0,$2^{14}$,$2^{15}$,$2^{16}$},
                ytick={0,8192,16384,32768,65536},
                yticklabels={0,$2^{13}$,$2^{14}$,$2^{15}$,$2^{16}$},
                xmin=-5000,
                xmax=70536,
                ymin=-2500,
                ymax=35268,
                grid=both,,
                minor grid style={opacity=0.3},
                major grid style={opacity=0.3},
                width=5.5cm,
                height=5.5cm,
                scaled x ticks=false,
                scaled y ticks=false,
            ]
            \addplot[MRC] table[x=N,y=I50] {figures/data/MRC.dat};
            \addplot[ORC] table[x=N,y=Imean] {figures/data/ORC.dat};
            \addplot[RS] table[x=N,y=Imean] {figures/data/RS.dat};
            \addplot[Harsha] table[x=N,y=Imean] {figures/data/Harsha.dat};
        \end{axis}
    \end{tikzpicture}
    \caption{Additional figures for the example of communicating categorical samples. \textit{Left:} The sample quality as a function of the coding cost, as in the main text but for a wider range of values. Note that samples of low quality (high $D_\text{TV}$) are rarely of interest. \textit{Middle:} The sample quality as a function of the maximum number of candidates available to an algorithm. \textit{Right:} The average number of candidates considered (that is, the number of iterations before termination) as a function of the maximum number of candidates.}
\end{figure}

\end{document}